\documentclass[twoside,11pt]{article}

\usepackage[preprint]{jmlr2e}

\usepackage{amsmath}
\usepackage{amssymb}
\usepackage{booktabs}       % professional-quality tables
\usepackage[scr=boondoxo]{mathalfa}
\usepackage{relsize}
\usepackage{multirow}
\usepackage{mathtools}
\usepackage{bbm}
\usepackage{colortbl}
\DeclarePairedDelimiter\ceil{\lceil}{\rceil}
\usepackage{algpseudocode}
\usepackage{enumitem}

\usepackage{caption,stackengine}
\algblock{ParFor}{EndParFor}
\algnewcommand{\LineComment}[1]{\State \(\triangleright\) #1}
\algnewcommand\algorithmicparfor{\textbf{parfor}}
\algnewcommand\algorithmicpardo{\textbf{do}}
\algnewcommand\algorithmicendparfor{\textbf{end\ parfor}}
\algrenewtext{ParFor}[1]{\algorithmicparfor\ #1\ \algorithmicpardo}
\algrenewtext{EndParFor}{\algorithmicendparfor}
\usepackage[ruled]{algorithm}
\newcommand{\bigcell}[2]{\begin{tabular}{@{}#1@{}}#2\end{tabular}}

\newcommand\scalemath[2]{\scalebox{#1}{\mbox{\ensuremath{\displaystyle #2}}}}
\ShortHeadings{Distributed Bayesian clustering}{Song and Wang and Dunson}
\firstpageno{1}

\savestack\mytable{
   \begin{tabular}{l|l}
Item index & Group \\ \hline
(3,18)     & 1     \\
(3,17)     & 2     \\
(3,16)     & 3     \\
(3,3)      & 4     \\
(3,2)      & 5     \\
(3,1)      & 6    
  \end{tabular}
}

\begin{document}

\title{Distributed Bayesian clustering using finite mixture of mixtures}

\author{\name Hanyu Song \email hanyu.song@duke.edu \\
       \addr  Department of Statistical Science\\
 Duke University\\
 Durham, NC 27708, USA
       \AND
       \name Yingjian Wang \email Yingjian.Wang@sas.com\\
       \addr SAS Institute Inc.\\
Cary, NC 27513, USA 
\AND
       \name David B.\ Dunson \email dunson@duke.edu \\
       \addr Department of Statistical Science\\
 Duke University\\
 Durham, NC 27708, USA}

\editor{Kevin Murphy and Bernhard Sch{\"o}lkopf}

\maketitle

\begin{abstract}%   <- trailing '%' for backward compatibility of .sty file
In many modern applications, there is interest in analyzing enormous data sets that cannot be easily moved across computers or loaded into memory on a single computer. In such settings, it is very common to be interested in clustering. Existing distributed clustering algorithms are mostly distance or density based without a likelihood specification, precluding the possibility of formal statistical inference. Model-based clustering allows statistical inference, yet research on distributed inference has emphasized nonparametric Bayesian mixture models over finite mixture models. To fill this gap, we introduce a nearly embarrassingly parallel algorithm for clustering under a Bayesian overfitted finite mixture of Gaussian mixtures, which we term distributed Bayesian clustering (DIB-C). DIB-C can flexibly accommodate data sets with various shapes (e.g. skewed or multi-modal). With data randomly partitioned and distributed, we first run Markov chain Monte Carlo in an embarrassingly parallel manner to obtain local clustering draws and then refine across workers for a final clustering estimate based on \emph{any} loss function on the space of partitions. DIB-C can also estimate cluster densities, quickly classify new subjects and provide a posterior predictive distribution. Both simulation studies and real data applications show superior performance of DIB-C in terms of robustness and computational efficiency.
\end{abstract}

\begin{keywords}
  Distributed algorithm, Model-based clustering, Bayesian methods, Markov chain Monte Carlo, Loss function
\end{keywords}

\section{Introduction}

Recent technological advances have greatly accelerated data collection processes, leading to explosively growing data sizes. These data sets are often too big to be stored on a single computer and too costly to move across computers. One common query to these data sets is cluster analysis, which seeks to group observations that are cohesive and separated from other groups. Large scale data sets from astronomy, flow cytometry and many other fields raise questions as to how to discover underlying clusters quickly while allowing for statistical inference.

To cluster large scale data sets, parallel and distributed clustering algorithms have been proposed. A common procedure underlying these algorithms is splitting the data into subsets and determining cluster assignments independently for each subset. Because clusters describe inherent relationships among all the data points, the independent local clustering must be carefully adjusted on the global scale via communication of local results. 

Most such algorithms are based on either distance or density without a likelihood specification; examples include density based distributed clustering \citep{Januzaj:2004:SDD:1053072.1053095}, K-Means with Map-Reduce (PKMeans) \citep{Zhaoetal2009} and co-clustering with Map-Reduce \citep{PapaSun2008}. These methods do not, in general,  have established statistical properties. An alternative method is model-based clustering, which considers the data as coming from a mixture distribution with different distributions for each cluster. Unlike the aforementioned methods, model-based clustering uses a soft assignment, where each data point has a probability of belonging to each cluster; it also allows density estimation and other statistical inference. 

One commonly used framework for model-based clustering is finite mixtures. Let $\mathcal{Y} = ({y}_1, \ldots, {y}_N)$, ${y}_i \in \mathbb{R}^d$ be a sample of size $N$. A finite mixture model assumes that ${y}_i\,(i = 1, \ldots, N)$ of dimension $d$ is generated from a finite mixture with $K$ exchangeable mixture components:
\begin{align}
f(y_i \mid \Theta, \eta) = \sum_{k = 1}^K \eta_k f_k(y_i\mid\theta_k),\quad  \Theta = (\theta_1, \ldots, \theta_K)\label{equ:finmix},
\end{align}
where $\eta_k$ is the weight associated with component $k$ satisfying $\sum_{k =1}^K \eta_k = 1$ and $f_k(y_i\mid\theta_k)$ is the component density specified by parameter $\theta_k$. With each component interpreted as a cluster, this model has been successfully applied to many areas, including agriculture, astronomy, bioinformatics, biology, economics, engineering, genetics, etc. 

%A Bayesian model puts priors on the model parameters (e.g. $\eta_k$ and $\theta_k$) enabling posterior inference. 
Finite mixture models often require a predetermined $K$, but the number of clusters is generally unknown. Even though one can fit multiple models with different $K$ and identify the best one based on model selection criteria (e.g. BIC), such a procedure can be time-consuming given a large data set and less appealing than the natural Bayesian approach, which is to treat the \emph{true} number of clusters $K_{true}$ as an unknown parameter to be estimated jointly with the component-specific parameters. Models that adopt this Bayesian approach include Bayesian nonparametric (BNP) mixture models and overfitted finite mixture models, where the number of clusters is automatically inferred from the observed data. 

BNP mixture models have been widely used for clustering in topic modelling \citep{Ge15distdpmm} and biomedical applications, because the model complexity adapts to the increasing amount of data.  In particular, the number of clusters grows with sample size.  Although this is a conceptually appealing property, in massive datasets this can lead to an enormous number of clusters.  This in turn creates an associated large computational burden and decreases interpretability and data simplification, two of the primary goals of clustering.  In addition, the inferred clusters may not represent actually distinct groups in the data, as extra clusters can arise as an artifact of BNP priors and due to inadequacies of typical kernels (e.g., Gaussian) in describing cluster shapes.  To address the computational problem, there is a rich and growing literature on scalable algorithms, using sequential approximations \citep{WangDunson11fastdpmm,Lin13BNPsequential,Tank15streamVIbnp} and parallelization.    

One common strategy underlying some parallel algorithms \citep{WilliamsonDubeyXing13parbnp,DubeyWilliamsonXing14parpityo, Ge15distdpmm} is exploiting conditional independence of cluster allocation given all other parameters to run Markov chain Monte Carlo (MCMC) in parallel. For example, \cite{WilliamsonDubeyXing13parbnp} proposed  parallel inference for Dirichet process (DP) mixture and hierarchical DP mixture models through re-parametrisation of a DP mixture model as a mixture of DPs. They further assume that each cluster only resides on one processor, which means that conditional on the processor allocations, the data points are distributed according to independent Dirichlet processes, facilitating parallel draws of the local cluster assignment. A global procedure to ensure that each cluster indeed resides on a single processor is run at every iteration and can require immense data transfer in a distributed system. This assumption can also lead to load imbalance if the cluster sizes are not uniform. Based on an alternative representation of a DP random measure, \cite{Ge15distdpmm} developed a slice sampler under the Map-Reduce framework for the same models, which is better suited for a distributed system. Summary statistics, instead of raw data, are transmitted from every mapper to a reducer for drawing global parameters at every iteration. Such communication, given the poor mixing and slow convergence of MCMC in the presence of latent variables, can be expensive.

%\cite{DubeyWilliamsonXing14parpityo} devised a similar parallel inference algorithm for Pitman-Yor mixture (PYM) models. To match the assumption that a cluster is local to a processor,  cluster reassignment of data points at each iteration Both algorithms are better suited for shared-memory systems, 
%In the cate \cite{WilliamsonDubeyXing13parbnp} proposed a shared-memory parallel inference algorithm for DP mixture (DPM) models using MCMC, in which they re-parametrise a DPM as a mixture of DPs and exploit the resulting conditional independence. This means that a cluster is assumed to only reside on one processor and cluster reassignment of data points at each iteration to match this assumption can require massive data transfer in a distributed-memory system. This assumption can also lead to load imbalance given very uneven cluster sizes. Taking a similar approach, \cite{DubeyWilliamsonXing14parpityo} devised a shared-memory parallel inference algorithm for Pitman-Yor mixture (PYM) models by expressing them as a finite Dirichlet mixture of PYM.

Other more efficient parallel algorithms focus on approximate inference under a BNP model; see, for example, distributed algorithms SNOB and SIGN by \cite{Zuanetti19largebnp} and \cite{Ni20parbnp} respectively. Instead of communicating at every iteration, they draw samples of cluster assignments locally, determine an optimal clustering estimate for each subset and then communicate sufficient statistics and clustering results for adjustment. 
%They continue to allocate the local clusters using MCMC, meaning that the clusters will never be split but possibly merged, and estimate an optimal clustering result based on these newly drawn clustering samples. 
%To obtain a final clustering estimate, SNOB clusters all the local clusters in one machine whereas SIGN iteratively performs the above adjustment procedures until the number of clusters is sufficiently small to be clustered in a single machine. 
A deficiency of SIGN is that if clusters are incorrectly merged at some iteration, then there is no hope of recovering the true clustering structure because they can never be split. In addition, both algorithms use a loss function multiple times at different stages to arrive at a final clustering estimate, raising the question as to whether the final clustering is actually a good approximation to the minimizer of the posterior expected loss.

Despite the proliferation of fast inference algorithms for BNP mixtures, there have been few similar advances for finite mixtures. Our view is that finite mixtures provide a more practically reasonable framework for clustering in massive datasets.  Most BNP approaches, including 
widely used DP mixtures and Pitman-Yor (PY) process mixtures, carry an implicit assumption that as the sample size goes to infinity, the number of clusters inevitably tends to infinity. For the DP mixtures 
$K_{true} \sim \alpha \log(N)$ \citep{korwar1973}, while for PY mixtures $K_{true} \sim N^{\beta}$ \citep{MillHarrison14,Orbanz2014}, where $\alpha$ and $\beta \in [0,1)$ are constants. This means the posterior of DP or PY mixtures fails to concentrate at the true number of components for data from a finite mixture.  

In contrast, overfitted finite mixtures apply a finite mixture model with the number of components $K$ intentionally set to be greater than the true number of clusters $K_{true}$. This approach is more useful for data with a moderate number of clusters that does not increase as the number of observations $N$ increases. Setting the prior on the mixture weights $\eta$ to be $\text{Dir}(e_0)$,  \cite{RouMen11} studied the asymptotic behavior of its posterior distribution: if $e_0 < d /2$, where $d$ is the dimension of the cluster-specific parameters $\theta_k$, the posterior expectation of the weights associated with empty clusters asymptotically converges to zero. Therefore the true number of clusters can be identified asymptotically despite the identifiability problems inherent in mixture models. 

A critical component of any mixture model specification is the choice of kernel. Although the Gaussian distribution is typically used, most data clusters in real world applications deviate from such a simple symmetric choice. Such misspecification can lead to identification of extraneous clusters as a means to improve model fit, which essentially destroys the interpretation of a mixture component as one cluster. A remedy is to instead model each cluster by a finite Gaussian mixture, a model that can accurately approximate a wide class of distributions \citep{Wallietal17} (henceforth MWFSG). Formally, this means each component distribution $f_k({y}_i \mid\theta_k)$ in \eqref{equ:finmix} is assumed to be a mixture of $L$ Gaussian subcomponents:
\begin{align}
f_k({y}_i \mid \theta_k) = \sum_{l = 1}^L \omega_{kl} p(y_i \mid \mu_{kl}, \Sigma_{kl}), \label{equ:subcomp}
\end{align}
where $\theta_k = \{\omega_{kl}, \mu_{kl}, \Sigma_{kl}\}_{l = 1}^L$ and $p(y_i \mid \mu_{kl}, \Sigma_{kl})$ are Gaussian densities used to approximate $f_{k}({y}_i \mid \theta_k)$. Setting $K$ in \eqref{equ:finmix} to be an upper bound of the number of clusters in the data yields a model called overfitted finite mixture of Gaussian mixtures. This model encounters identifiability issues due to exchangeability of all the subcomponents; fortunately MWFSG developed prior specification for this model that encourages subcomponents within a cluster to be close and clusters to be spread out, minimizing these issues. 

%Although Bayesian mixture models are appealing in providing uncertainty quantification in clustering and associated inferences, a key disadvantage is the large computational cost, particularly as the sample size grows.  Joint posterior distributions typically have an intractable analytic form, and one typically relies on Markov chain Monte Carlo (MCMC) sampling to generate draws from the posterior. Conventional MCMC algorithms are too slow to be practically useful in massive data applications. Their inherently serial nature also poses challenges in constructing a parallel counterpart. \emph{Embarrassingly parallel} (EP) MCMC algorithms have been recently developed to deal with this problem  (\citealp{scott:2016:epmcmc,neiswanger:2014:epmcmc,minsker:2014:epmcmc,srivastava:2015:epmcmc,wang:2015:epmcmc, Lietal16}).  The idea is to break the data into subsets, run MCMC in parallel for each subset and then combine the results.

%Inspired by embarrassingly parallel MCMC
In this article we propose a distributed Bayesian inference method based on (overfitted) finite mixture of Gaussian mixtures for clustering, which we refer to as DIB-C. 
Our main contributions lie in developing a decision theoretic approach to identifying a reliable clustering estimate while minimizing data transmission between the master and workers, a key consideration in distributed computing. We adopt a strategy used in SNOB and SIGN, which is to produce MCMC samples of local clustering assignments in an embarrassingly parallel manner to minimize data communication. Unlike SNOB or SIGN, our adjustment to local clusters permits both cluster merging and splitting, and our clustering estimate is more reliable as we only apply a loss function \emph{once} in the entire framework to the samples of adjusted local clusterings. These steps are enabled by one of the simplest parallel programming paradigm, \emph{master-worker}, and the communication of summary statistics at some iterations between the master and workers. 
%use a decision-theoretic approach and only apply the loss function \emph{once} in the entire framework.
%In addition, to identify a more reliable clustering estimate, we use a decision-theoretic approach that produces a final clustering estimate that minimizes a loss function (as opposed to other algorithms where local cluster assignments, rather than global assignments, are selected to minimize loss functions).
%For scalability, we produce samples of cluster assignments in each partition of the data and combine them for an approximate sample of global cluster allocations. 

In addition to clear computational gains, DIB-C exhibits superior clustering performance in comparison to its non-distributed counterpart. In addition, DIB-C can accommodate any loss function on the space of partitions for cluster estimation and enables density estimation, quick classification of new subjects, sampling from the posterior predictive distribution and uncertainty quantification of cluster-specific parameters (such as cluster centers). As a side effect, DIB-C also works for semi-supervised clustering: when the true number of clusters is greater than that represented in the labeled data, DIB-C can automatically determine the number of clusters via the use of an overfitted finite mixture model.

The rest of the paper is organized as follows. In Section \ref{sec:model}, we review the model and prior specification of finite mixture of mixtures. In Section \ref{sec:method}, we describe the DIB-C framework. Section \ref{sec:experiment} presents extensive experimental results to illustrate the performance of our framework.

\section{Model Specification: Finite Mixture of Gaussian Mixtures\label{sec:model}}
 Clusters are groups of data points that are cohesive and connected within a group but are separated from other groups. Assume $\mathcal{Y}$ can be partitioned into $R$ non-overlapping subsets, with subset $r$, denoted by $\mathcal{Y}_r$, residing on worker $r$ $(r = 1, \ldots, R)$.  
 %In this article, we focus on identifying convexly shaped clusters, of which the cluster centers are a good representative, in a distributed manner. 
 The model formulation, as defined by \eqref{equ:finmix} and \eqref{equ:subcomp}, leads to a hierarchy: the upper level \eqref{equ:finmix} captures a {heterogeneous} population with $K$ different clusters and each cluster corresponds to a mixture component; the lower level \eqref{equ:subcomp} approximates each cluster distribution via a mixture of Gaussian densities $p(y_i\mid \mu_{kl}, \Sigma_{kl})$. To distinguish the upper and lower level components, we adopt the convention in MWFSG to call $f_{k}({y}_i\mid\theta_k)$ \emph{cluster distribution $k$} and $p(y_i\mid \mu_{kl}, \Sigma_{kl})$ \emph{subcomponent distribution} $l$ in cluster $k$. 

Combining \eqref{equ:finmix} and \eqref{equ:subcomp} provides an alternative expression of the likelihood of ${y}_i$:\begin{align}f({y}_i \mid \Theta, \eta) = \sum_{k = 1}^K \sum_{l = 1}^{L} \upsilon_{kl} p(y_i \mid \mu_{kl}, \Sigma_{kl}),\label{equ:mod2}\end{align}where $\upsilon_{kl} = \eta_k \omega_{kl}$. \eqref{equ:mod2} is invariant to  permutations of the $K \cdot L$ subcomponents: exchanging subcomponents between clusters does not alter the likelihood, but goes contrary to the general characterization of data clusters, which is a densely connected cloud of data points far away from other densely connected ones. To incorporate such structure, MWFSG proposed a two-level hierarchical prior that repulses the cluster centers and attracts subcomponent means towards the cluster centers. Additionally, since cluster structure is invariant to the ordering of both clusters and subcomponents within a cluster, symmetric priors should be used for clusters on the upper and lower level, respectively, to ensure exchangeability. 

Let $\varphi_0 = (e_0, d_0, c_0, g_0, \mathbf{G}_0, \mathbf{B}_0, \mathbf{m}_0, \mathbf{M}_0,\nu)$ be a set of fixed hyper-parameters. The priors at the cluster level are specified so that the $K$ clusters are exchangeable: 
\begin{align}
p(\eta, \theta_1, \ldots, \theta_K \mid \varphi_0) = p(\eta \mid e_0)\prod_{k = 1}^K p(\theta_k \mid \varphi_0),\label{equ:clusterprior}
\end{align}
where $\eta\mid e_0 \sim \text{Dir}_K(e_0)$, and $\theta_k \mid \varphi_0$ are independent and identically distributed a priori. 
% \ref{equ:clusterprior} is 
Within each cluster $k$, the prior distribution can be factored as: \begin{align}p(\theta_k \mid\varphi_0) = p(w_k\mid d_0) p(\mu_{k1}, \mu_{k2}, \ldots, \mu_{kL}\mid\mathbf{B}_0, \mathbf{m}_0, \mathbf{M}_0, \nu)p(\Sigma_{k1}, \Sigma_{k2}, \ldots, \Sigma_{kL}\mid c_0, g_0, \mathbf{G}_0),\label{equ:subcompprior}\end{align} where
$\omega_k \mid d_0 \overset{iid}{\sim} \text{Dir}_L(d_0)$, $\mu_{k1}, \ldots, \mu_{kL}$ are independently distributed conditional on $\mathbf{B}_0, \mathbf{m}_0,\allowbreak \mathbf{M}_0,\nu$, 
and $\Sigma_{k1}, \ldots, \Sigma_{kL}$ are independent conditional on $c_0, g_0, \mathbf{G}_0$.

To create the conditional independence in \eqref{equ:clusterprior} and \eqref{equ:subcompprior}, MWFSG formulated hierarchical ``random effects'' priors: first the cluster-specific parameters $(\mathbf{C}_{0k}, \mathbf{b}_{0k}, \Lambda_k)$  are drawn from the same set of distributions and then, conditional on these, the subcomponent-specific parameters $(\mu_{kl}, \Sigma_{kl})_{l = 1}^L$ within cluster $k$ are drawn from another set of distributions for all $k$.

Specifically, cluster-specific parameters  $(\mathbf{C}_{0k}, \mathbf{b}_{0k})$ and $\Lambda_k = diag(\lambda_{k1}, \ldots, \lambda_{kd})$, $k = 1, \ldots, K$ are drawn from: 
%Conditional on the fixed upper level hyperparameters $(g_0, \mathbf{G}_0, \mathbf{m}_0,\mathbf{M}_0, \nu)$, cluster-specific random hyperparameters $(\mathbf{C}_{0k}, \mathbf{b}_{0k})$ and $\Lambda_k = diag(\lambda_{k1}, \ldots, \lambda_{kd})$, $k = 1, \ldots, K$ have priors as follows:
\begin{align*}
\mathbf{C_{0k}} \mid g_0, \mathbf{G}_0 &\sim \text{W}_d(g_0, \mathbf{G}_0),\\\mathbf{b}_{0k} \mid \mathbf{m}_0,\mathbf{M}_0 &\sim \text{N}_d(\mathbf{m}_0, \mathbf{M}_0), \\ (\lambda_{k1}, \ldots, \lambda_{kd})\mid \nu &\sim \text{Ga}(\nu, \nu),  \end{align*}
where $\mathbf{m}_0$ is the overall data center. Cluster centers $\mathbf{b}_{0k}$ are generated around $\mathbf{m}_0$ with $\mathbf{M}_0$ controlling the shrinkage of $\mathbf{b}_{0k}$ towards $\mathbf{m}_0$. MWFSG set $\mathbf{M}_0 \gg \mathbf{S}_\mathcal{Y}$, where $\mathbf{S}_\mathcal{Y}$ is the sample covariance of all data so that cluster centers lie relatively far from each other.

Conditional on the cluster-specific random hyperparameters $(\mathbf{C}_{0k}, \mathbf{b}_{0k}, \Lambda_k)$ and the fixed lower level hyperparameters $(\mathbf{B}_0, c_0)$, the $L$ subcomponent means $\mu_{kl}$ and covariance matrices $\Sigma_{kl}$ are drawn independently for all $l = 1, \ldots, L$: 
\begin{align}
\mu_{kl} \mid \mathbf{B}_0, \mathbf{b}_{0k}, \Lambda_k &\sim \text{N}_d(\mathbf{b}_{0k}, \sqrt{\lambda_k} \mathbf{B}_0 \sqrt{\lambda_k}), \label{equ:priorsubcomp}\\ 
\Sigma_{kl}^{-1} \mid c_0, \mathbf{C}_{0k} &\sim \text{W}_d(c_0, \mathbf{C}_{0k}).\end{align}
$\mu_{kl}$ ($l = 1,\ldots, L$) should be close to the cluster center $\mathbf{b}_{0k}$ to ensure no gaps among subcomponents and $\Sigma_{kl}$ ($l = 1, \ldots, L$) should be diffuse so that the boundary or outlier points can be well fitted. Therefore, we need $\mathbf{B}_0$ to impose strong shrinkage of $\mu_{kl}$'s toward  $\mathbf{b}_{0k}$ and $c_0$ to be small to induce large variances in $\Sigma_{kl}$'s while permitting large variation of $\Sigma_{kl}$'s.

To elicit the prior, MWFSG decomposes the variation of an observation ${y}$ into three sources: 
\begin{align*}\text{cov}({y}) &= \underbrace{\sum_{k= 1}^K\eta_k \Sigma_k}_{\text{within cluster variation}} + \underbrace{\sum_{k = 1}^K \eta_k \mu_k \mu_k^\prime - \mathbf{\mu}\mathbf{\mu}^\prime}_{\text{between cluster variation}}\\
&= \underbrace{\sum_{k = 1}^K \eta_k \sum_{l = 1}^L \omega_{kl} \Sigma_{kl}}_{\text{within subcomponent variation}} + \underbrace{\sum_{k = 1}^K\left(\sum_{l = 1}^L\omega_{kl} \mu_{kl} \mu_{kl}^\prime - \mu_k\mu_k^\prime\right)}_{\text{within cluster, between subcomponent variation}}\\& + \underbrace{\sum_{k = 1}^K \eta_k \mu_k\mu_k^\prime - \mu \mu^\prime}_{\text{between cluster variation}} \\ 
&:= (1 - \phi_W)(1 -\phi_B)\text{cov}({y}) + \phi_W(1 - \phi_B)\text{cov}({{y}}) + \phi_B\text{cov}({y}),
\end{align*} 
where $\phi_B$ is the proportion of variability explained by the cluster centers $\mu_k$ around the overall mean $\mu = \sum_{k = 1}^K \eta_k \mu_k$, and $\phi_W$ is the proportion of total variation explained by subcomponent means $\mu_{kl}$ $(l = 1, \ldots, L)$ around cluster center $\mu_k = \sum_{l = 1}^L \omega_{wl}\mu_{kl}$, for all $k$. By setting $\phi_B = 0.5$ and $\phi_W = 0.1$, we balance the variation from the three sources and can elicit prior parameters accordingly. 

In this article, the MCMC sampling scheme is derived based on the above specified prior distributions, but our DIB-C framework is broadly applicable to any prior specification for a finite mixture of mixtures.

\section{DIB-C Framework\label{sec:method}}

The general idea of the DIB-C framework is to produce approximate MCMC samples of cluster assignments under the assumption of a finite mixture of mixtures model in a distributed computing paradigm. Specifically, we sample cluster assignments in each partition of the data and combine them for approximate samples of global cluster allocations.

Because clusters describe inherent relationships among all the data points, the independent clustering performed on each subset must be carefully merged. To minimize data transmission, we communicate only sufficient statistics between the master and workers to refine local cluster assignments across all partitions of the data in order to produce reliable global cluster allocations.

\begin{figure}
    %centering
    \includegraphics[width = 0.95 \textwidth]{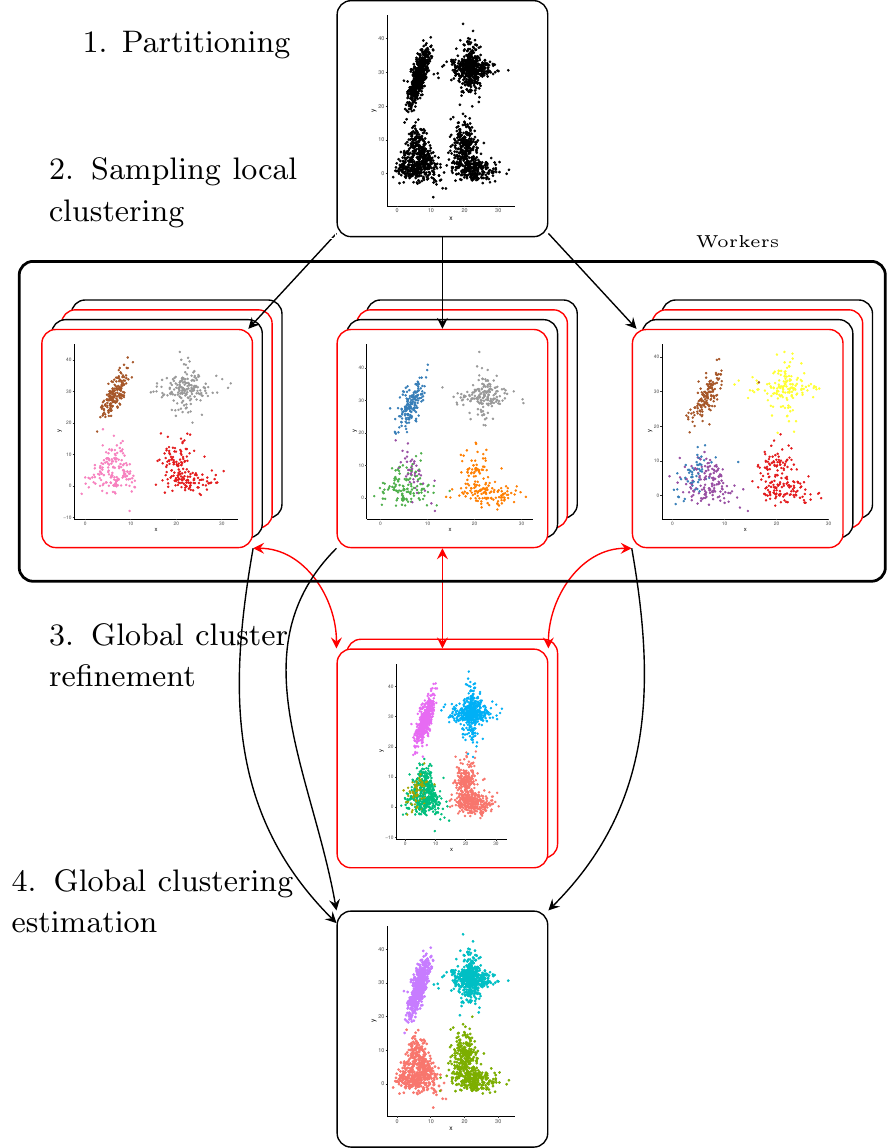}
    \caption{Algorithm flowchart of DIB-C. In step 2, the copies of frames refer to samples of local clustering, with the red ones representing those that are adjusted in step 3. Based on the adjusted samples of local clustering, a global clustering is estimated in step 4. Step 5 Sampling model parameters is excluded due to space constraint.}
    \label{fig:alg_diag}
\end{figure}
Based on the above discussion, we propose a DIB-C framework that consists of five steps: a. \emph{partitioning}: randomly partitioning the data $\mathcal{Y}$ and distributing them over $R$ workers; b. \emph{sampling local clustering}: running embarrassingly parallel MCMC to obtain samples of subcomponent and cluster assignments on each partition of the data based on a finite mixture of mixtures model; c. \emph{global cluster refinement}: refining some samples of local clustering via sufficient statistics from each subset; d. \emph{global clustering estimation}: from refined local cluster assignments, identifying a global clustering estimate that minimizes the expected posterior loss defined on the space of partitions; and e. \emph{sampling model parameters}: conditional on the optimal global cluster and subcomponent estimate, drawing model parameters from MCMC and performing inference. See Figure \ref{fig:alg_diag} for a diagram of the procedures.

Global cluster refinement (step c), global clustering estimation (step d) and sampling model parameters (step e) incur a small amount of data transmission between the master and workers. In the rest of this section, we introduce the notations used throughout the paper and elaborate on steps b, c, d and e. 

\subsection{Notation}
\begin{table}[]
\caption{Selected notation and descriptions}
\label{tab:notation}
\begin{tabular}{cl}
\hline
Notation                                                                                                                                        & Description                                                                                                                                                         \\ \hline
$\mathcal{Y} = \{y_1, \ldots,y_N\}$                                                                                                                                   & Sample of size $N$                                                                                                                                                            \\
$K$                                                                                                                                           & A pre-determined upper bound of number of clusters                                                                                                                  \\
$L$                                                                                                                                             & Number of subcomponents per cluster                                                                                                                                 \\
$r$, $r = 1, \ldots, R$                                                                                                                         & Subset or worker index                                                                                                                                              \\
Subscript $ri$                                                                                                                                  & Data point $i$ in subset $r$                                                                                                                                        \\
$\mathcal{Y} = \{\mathcal{Y}_1, \ldots, \mathcal{Y}_R\}$                                                                                        & The sample partitioned into $R$ non-overlapping subsets                                                                                                          \\
$\mathcal{Y}_r = \{y_{r1}, y_{r2}, \ldots, y_{rn_r}\}$                                                                                          & Data subset $r$ of size $n_r$, with $y_{ri}$ being data point $i$                                                                                                   \\
$\mathbf{c} = \{\mathbf{c}_1, \ldots, \mathbf{c}_R\}$                                                                                           & Latent cluster allocations                                                                                                                                          \\
$\mathbf{s} = \{\mathbf{s}_1, \ldots, \mathbf{s}_R\}$                                                                                           & Latent subcomponent allocations                                                                                                                                     \\
$\mathbf{c}_r= \{{c}_{r1}, \ldots, {c}_{rn_r}\}$                                                                                                & \begin{tabular}[c]{@{}l@{}}Latent cluster allocations for subset $r$, where\\ ${c}_{ri} = j$ indicates data point $i$ belongs to cluster $j$\end{tabular}           \\
$\mathbf{s}_r = \left(s_{r1}, \ldots, s_{r n_r}\right)$                                                                                         & \begin{tabular}[c]{@{}l@{}}Latent subcomponent allocations for subset $r$, where\\ ${s}_{ri} = j$ indicates data point $i$ belongs to subcomponent $j$\end{tabular} \\ \hline
\multicolumn{2}{c}{(Global Cluster Refinement)}                                                                                                                                                                                                                                                                       \\ \hline
$\tilde{\mathbf{c}} = \{\tilde{\mathbf{c}}_1, \ldots, \tilde{\mathbf{c}}_R\}$                                                                   & Updated cluster allocations                                                                                                                                         \\
$\tilde{\mathbf{s}} = \{\tilde{\mathbf{s}}_1, \ldots, \tilde{\mathbf{s}}_R\}$                                                                   & Updated subcomponent allocations                                                                                                                                    \\
Item                                                                                                                                            & Non-empty subcomponent                                                                                                                                              \\
Item $(r, j)$                                                                                                                                   & \begin{tabular}[c]{@{}l@{}}First item indexing set; $r$ is the worker/subset index \\ and $j$ is the item index within the worker\end{tabular}                      \\
Item $(r, L(k - 1) + l)$                                                                                                                        & \begin{tabular}[c]{@{}l@{}}Item index for the $l$th subcomponent \\ in cluster $k$ in subset $r$\end{tabular}                                                       \\
$\mathscr{L}_r$                                                                                                                                 & Item counts in subset $r$                                                                                                                                           \\
$B = \sum_{r = 1}^R\mathscr{L}_r$                                                                                                               & Total number of items across subsets                                                                                                                                \\
\begin{tabular}[c]{@{}c@{}}item $b$, \\ $b = 1, \ldots, B$\end{tabular}                                                                         & \begin{tabular}[c]{@{}l@{}}Second item indexing set \\ created by an ordering induced by the workers\end{tabular}                                                   \\
$\mathscr{z}_{j} \in \{1, \ldots, H\}$ and $\tilde{\mathscr{z}}_{j}$                                                                            & Group allocation and updated label of item $j$ respectively                                                                                                         \\
$\mathscr{r}_j$                                                                                                                                 & The worker where item $j$ resides                                                                                                                                   \\
$\mathscr{r}^{*}$                                                                                                                               & The reference subset                                                                                                                                                \\
$\mathscr{I}_j$                                                                                                                                 & The set of data indices in item $j$                                                                                                                                 \\
$\mathscr{y}_{j}$, $\mathscr{n}_j$,$\bar{\mathscr{y}}_j$ and ${\mathscr{S}}_j$                                                                  & \begin{tabular}[c]{@{}l@{}}Data, sample size, first and second moment\\ associated with item $j$ respectively\end{tabular}                                          \\
\begin{tabular}[c]{@{}c@{}}$\mathscr{Y}_h$, $\mathscr{N}_{h}$, $\bar{\mathscr{Y}}_{h}$ and $\mathscr{S}_{h}$,\\ $h = 1, \ldots, H$\end{tabular} & \begin{tabular}[c]{@{}l@{}}Data, sample size, first and second moment \\ associated with group $h$ respectively\end{tabular}                                        \\
Subscript ${h \setminus b}$                                                                                                                     & Group $h$ without taking item $b$ into account                                                                                                                      \\ \hline
\multicolumn{2}{c}{(Global Clustering Estimation)}                                                                                                                                                                                                                                                                       \\ \hline
$C_i$                                                                                                                                           & The set of data indices in cluster $i$ of the true clustering                                                                                                       \\
$\widehat{C}_j$                                                                                                                                 & The set of data indices in cluster $j$ of clustering candidate $\hat{\mathbf{c}}$                                                                                   \\
$N_{i+}$                                                                                                                                        & $\vert C_i \vert$ under the true clustering                                                                                                                         \\
$N_{+j}$                                                                                                                                        & $\vert C_j \vert$ under clustering candidate $\hat{\mathbf{c}}$                                                                                                     \\
$N_{ij} = |{C}_i \bigcap \widehat{C}_j|$                                                                                                        & The number of data points in both $C_i$ and $\widehat{C}_j$                                                                                                         \\
$\mathscr{T}$                                                                                                                                   & The set of iterations refined in global cluster refinement                                                                                                          \\
$\mathscr{M} \subset \mathscr{T}$                                                                                                               & The set of iterations associated with the clustering candidates                                                                                                     \\
Superscript $(t)$ or $(t, t^\prime)$                                                                                                            & Iteration $t$ or iteration $t$ and $t^\prime$                                                                                                                       \\ \hline
\end{tabular}
\end{table}
We represent any quantity associated with subset or worker $r$ by adding a subscript $r$, $r = 1, \ldots, R$. Let the sample size of data subset $\mathcal{Y}_r$ be $n_r$, and $\mathcal{Y}_r  = \{y_{r1}, y_{r2}, \ldots, y_{rn_r}\}$. Assume subset $\mathcal{Y}_r$ is distributed to and processed on worker $r$, $r = 1, \ldots, R$.
As discussed in Section \ref{sec:model}, the finite mixture of mixtures model induces both latent cluster and subcomponent allocations of the data. Let a vector of latent cluster allocations be $\mathbf{c} = \left(c_1, \ldots, c_N\right)$, where $c_i = j$ indicates that data point $i$ belongs to cluster $j$.  Similarly let a vector of latent subcomponent allocations be $\mathbf{s}$, where $s_i = j$ indicates that data point $i$ belongs to subcomponent $j$.

%Alternatively, a latent cluster (resp. subcomponent) allocation can be represented by $\mathbf{c} = \left(c_1, \ldots, c_N\right)$ (resp. $\mathbf{s} = \left(s_1, \ldots, s_N\right)$), where $c_i = j$ (resp. $s_i = j$) indicates that data point $i$ belongs to cluster $j$ (resp. subcomponent $j$). 
In a distributed computing paradigm, we can also express $\mathbf{c}$ (resp. $\mathbf{s}$) as $\{\mathbf{c}_1, \ldots, \mathbf{c}_R\}$ (resp. $\{\mathbf{s}_1, \ldots, \mathbf{s}_R\}$), where $\mathbf{c}_r$ (resp. $\mathbf{s}_r$) represents a vector of latent cluster (resp. subcomponent) allocations for $\mathcal{Y}_r$. Both $\mathbf{s}$ and $\mathbf{c}$ are updated during global cluster refinement. Let the updated cluster allocations be $\tilde{\mathbf{c}} := \{\tilde{\mathbf{c}}_1, \ldots, \tilde{\mathbf{c}}_R\}$ (resp. $\tilde{\mathbf{s}} := \{\tilde{\mathbf{s}}_1, \ldots, \tilde{\mathbf{s}}_R\}$), where $\tilde{\mathbf{c}}_r$ (resp. $\tilde{\mathbf{s}}_r$) represents the updated cluster allocations based on $\mathbf{c}_r$ (resp. $\mathbf{s}_r$). Additional notation is introduced in the relevant sections. See Table \ref{tab:notation} for a collection of important notation and descriptions used in this article. 

%We also use subscript $\setminus{\cdot}$ to represent the associated quantity without taking $\cdot$ into account. 
\subsection{Sampling Local Clustering \label{sec:pie}}
Let the overall posterior density given $\mathcal{Y}$ and the $r$th subset posterior density given $\mathcal{Y}_r$, $r = 1, \ldots, R$ be
\begin{align}
p(\Theta, \eta \mid \mathcal{Y})  &= \frac{\{\prod_{r = 1}^R\prod_{i = 1}^{n_r} f(y_{ri} \mid \Theta, \eta)\}p(\Theta, \eta)}{\int \int \{\prod_{r = 1}^R\prod_{i = 1}^{n_r}f(y_{ri} \mid \Theta, \eta)\}p(\Theta, \eta)d\Theta d\eta} \quad\text{and}\nonumber\\ 
p_r(\Theta, \eta \mid \mathcal{Y}_r) &= \frac{\{\prod_{i = 1}^{n_r} f(y_{ri} \mid \Theta, \eta)\}p(\Theta, \eta)}{\int \int \{\prod_{i = 1}^{n_r} f(y_{ri} \mid \Theta, \eta)\}p(\Theta, \eta)d\Theta d \eta}\label{equ:subsetpost},
\end{align} respectively. In our algorithm, we run MCMC on $R$ workers in parallel based on \eqref{equ:subsetpost}, producing draws from each subset posterior $p_r(\Theta, \eta \mid \mathcal{Y}_r)$, $r = 1, \ldots, R$. 

To produce draws from $p_r(\Theta, \eta \mid \mathcal{Y}_r)$, $r = 1, \ldots, R$ (or $p(\Theta, \eta \mid \mathcal{Y})$ in a serial algorithm), we can run a block conditional Gibbs sampler with data augmentation. The sampler alternates between imputing cluster allocations and updating parameters specific to each cluster, with the latter step alternating between imputing subcomponent allocations and updating subcomponent-specific parameters.  See Appendix \ref{appendix:mcmc} for a detailed sampling scheme. One can, however, adopt any other sampling scheme to improve mixing (such as a collapsed Gibbs sampler) as long as samples of cluster and subcomponent allocations are produced.

\subsection{Global Cluster Refinement \label{sec:refine}}
\begin{figure}
  \centering
	\includegraphics[width =  \textwidth]{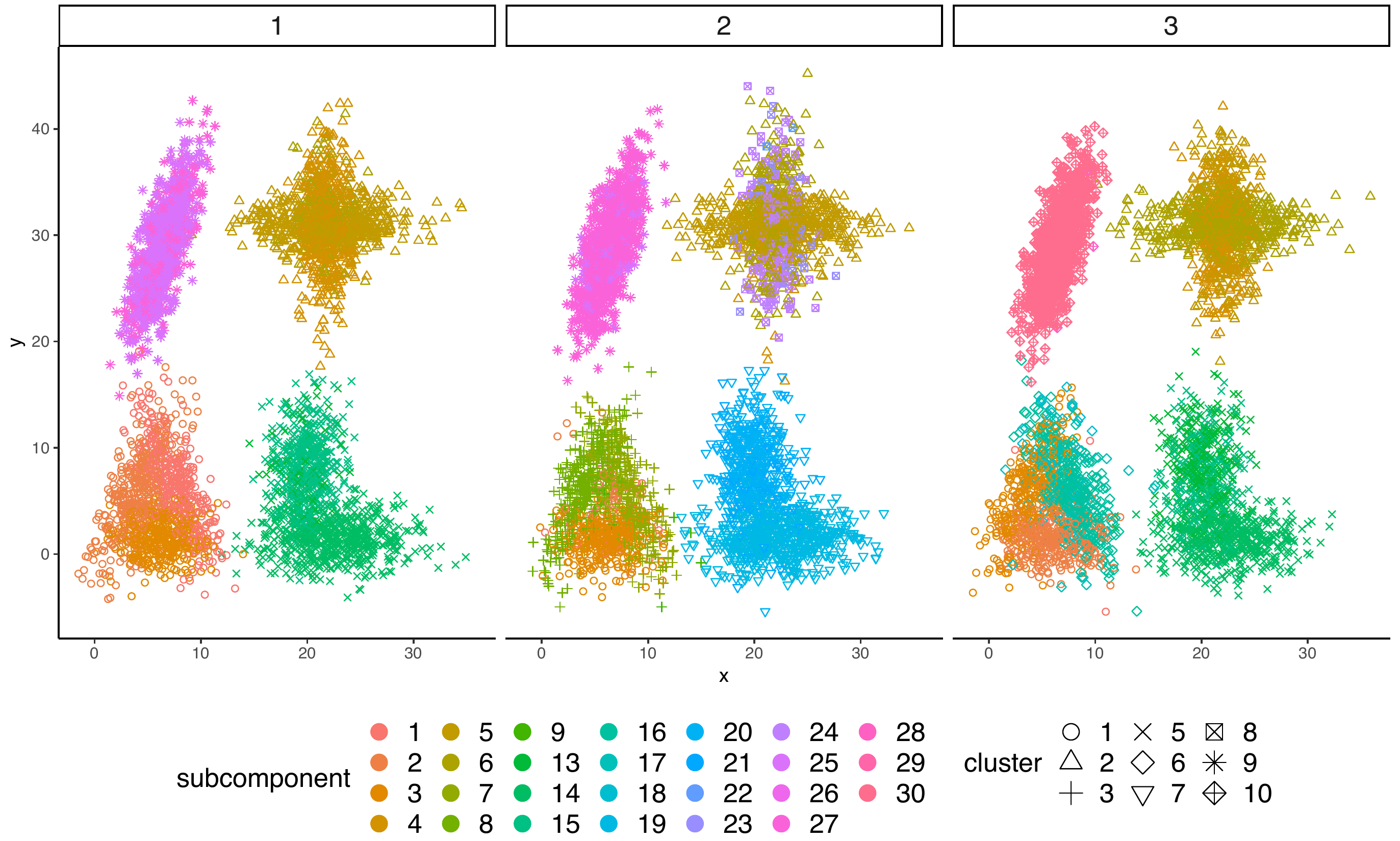}
	%\caption{A sample of local clusters (the first row) and subcomponent (the second row) allocation in subset 1, 2 and 3 when the data set is partitioned and run on 4 workers. $6 \%$ randomly selected data points are plotted.  \label{fig:refine_illus1}}
	\caption{A sample of local cluster (top) and subcomponent (bottom) allocations before refinement for subset 1, 2 and 3 when the data set is partitioned and distributed to 4 workers. $6 \%$ randomly selected data points are plotted.  \label{fig:refine_illus1}}
\end{figure}

The clustering $\mathbf{c} = \{\mathbf{c}_1, \ldots, \mathbf{c}_R\}$, generated from naively combining the clustering allocations from the subsets, does not, in general, mimic a sample of cluster allocations from $p(\Theta, \eta \mid \mathcal{Y})$ for two reasons. First, the cluster labeling can vary across workers. Figure \ref{fig:refine_illus1} shows a sample of local cluster allocations from three different workers when the data set is partitioned and distributed to 4 workers: cluster 9 in subset 1 corresponds to cluster 10 in subset 3. Second, the clustering structure could vary across subsets. For example, a single cluster in one subset (e.g. cluster 1 in subset 1 in Figure \ref{fig:refine_illus1}) can correspond to several smaller ones (e.g. cluster 1 and 3 in subset 2) in another subset. Even worse, a cluster in one subset corresponds to a significant number of, but not all, data points from multiple clusters in another subset. Therefore, an algorithm to adjust samples of local cluster assignments, particularly enabling merging and splitting clusters for handling the second issue, is essential.

\begin{figure}
    \centering
    \includegraphics[width = \textwidth]{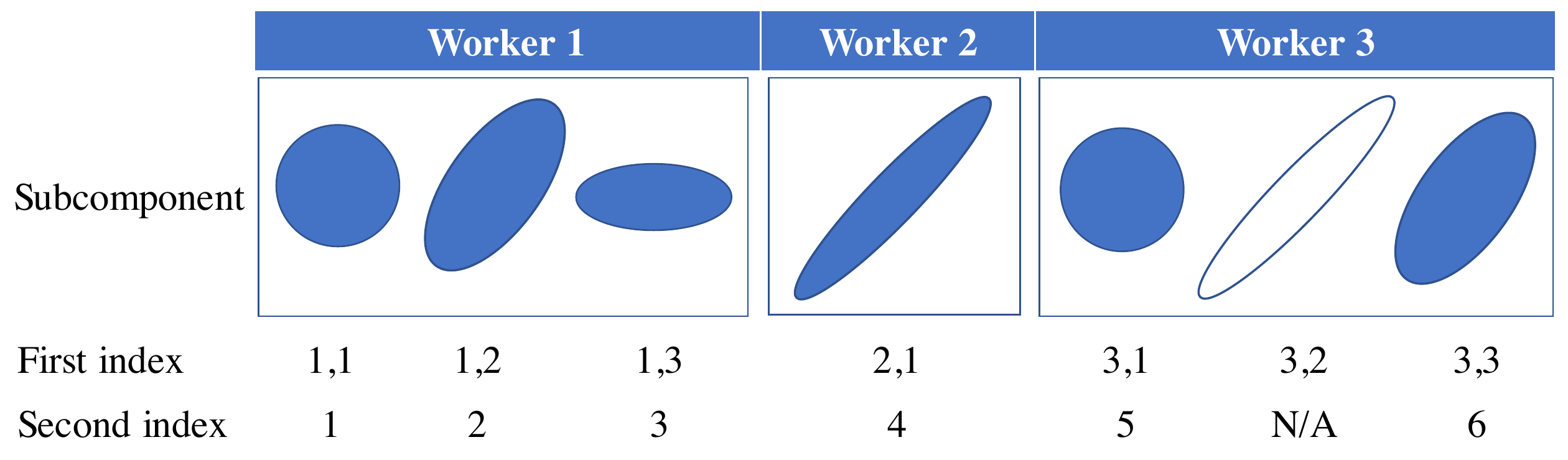}
    \caption{Representation of items via the two indexing systems.  Blue and white represent non-empty and empty subcomponents respectively.}
    \label{fig:item_idx}
\end{figure}
%\begin{figure}
%\centering
%	\includegraphics[width =0.65 \textwidth]{figures/method_illustration/syndat_afteralign_subcomp120k_nworkers4_run1_iter450_frac6.pdf}	
%	\includegraphics[width = 0.65 \textwidth]{figures/method_illustration/syndat_afteralign_clust120k_nworkers4_run1_iter450_frac6.pdf}	
%	\caption{A sample of local cluster (top) and subcomponent (bottom) allocations for subset 1, 2 and 3 after applying the global clustering refinement algorithm to the outputs shown in Figure \ref{fig:refine_illus1}. $6 \%$ randomly selected data points are plotted. \label{fig:refine_illus2}}
	%	\caption{A sample of local subcomponent allocations before (top) and after refinement (bottom) for subset 1, 2 and 3 when the data set is partitioned and distributed to 4 workers. $6 \%$ randomly selected data points are plotted.  \label{fig:refine_illus2}}
%\end{figure}
	
%\cite{Ni20parbnp} proposed a multi-step recursive approach to cluster merging. They \emph{freeze} the local clusters, meaning that the observations within each cluster will never be split but possibly merged in the subsequent steps, and randomly divide the frozen clusters into $R^\prime$ workers, where $R^\prime < R$, for further clustering based on the same BNP model. They iteratively perform the above procedures until the number of frozen clusters is sufficiently small to be clustered in a single node.

Inspired by \cite{Ni20parbnp}, we propose a simple and communication efficient algorithm that permits both cluster merging and splitting. Recall \cite{Ni20parbnp}'s multi-step recursive approach to cluster merging, in which they recursively \emph{freeze} the local clusters, meaning that the observations within each cluster will never be split but possibly merged in the subsequent steps.
%and randomly divide the frozen clusters into fewer workers for further clustering based on the same BNP model. 

In contrast, we freeze observations at the subcomponent level, rather than at the cluster level. One big advantage is that heavily overlapping subcomponents provide a natural solution: we can merge or group the frozen local subcomponents across subsets based on their degrees of overlap and map the updated subcomponent labels in an appropriate way to the cluster level, which results in automatic merges or splits of the clusters. Such a joint grouping scheme also ensures that a unified set of subcomponent labels is applied across subsets. Another advantage, since each subcomponent can be described by a Gaussian distribution, is that the natural model to enable such \emph{joint} grouping is simply a finite mixture of Gaussians: if the distributions of any two subcomponents can be well approximated by a single Gaussian kernel, the subcomponents are likely to be grouped together. In Figure \ref{fig:refine_illus1}, for example, subcomponent 1, 7 and 16 from subset 1, 2 and 3, respectively, are likely to be assigned to the same group due to their high overlap. 

This process depends on the data only through some summary statistics of the \emph{non-empty subcomponents}, which we refer to as \emph{items}, using the word ``item" to distinguish these objects from single observations. Denote the associated quantities of an item, including the group allocation, updated label, data, set of data indices, sample size, first and second moment and the worker to which it belongs by $\mathscr{z}_{\cdot}$, $\tilde{\mathscr{z}}_{\cdot}$, $\mathscr{y}_{\cdot}$, $\mathscr{I}_\cdot$, $\mathscr{n}_\cdot$, $\bar{\mathscr{y}}_\cdot$, ${\mathscr{S}}_\cdot$ and $\mathscr{r}_\cdot$ respectively, where $\cdot$ represents the index of the underlying item. The group allocation and the updated label are two slightly different concepts, as we will explain later.

For ease of demonstration, we herein introduce two different indexing systems for these items.
The first is represented by $(r, j),$ where $r$ is the worker/subset index and $j$ is the item index within the worker. Since each cluster contains $L$ subcomponents, we index the $l$th subcomponent in cluster $k$ of subset $r$ by $L(k -1)  + l$; if this subcomponent is non-empty, its item index is given by $L(k -1)  + l.$ 

The second indexing method is represented by the item index created by an ordering induced by the ordering of the workers. Let the item counts in subset $r$ be $\mathscr{L}_r$ and the total number of items across all subsets be $B$; we have $B = \sum_{r= 1}^R \mathscr{L}_r.$ Hence, the set of items can be naturally indexed by $\{1, 2, \ldots, B\}.$ Using this notation, for example, the worker where item $b$ resides can be expressed as $\mathscr{r}_b = \{r: \mathscr{y}_b \subset \mathcal{Y}_r, r = 1,\ldots, R\}.$ See Figure \ref{fig:item_idx} for an intuitive illustration of these two indexing systems.
%Denote the total item counts in subset $r$ by $\mathscr{L}_r$.
%For example, $B = 6$ for the example in Figure \ref{fig:item_idx}. 

The finite mixture of Gaussian we employ to group the items assumes that the density of $\mathscr{y}_b$ is given as follows:
 \begin{align}
 f({\mathscr{y}}_b \mid \Xi, \tau) = \sum_{h = 1}^{H}\tau_{h} \text{N}_h(\mathscr{y}_b \mid \xi_h), \label{equ:group_mod}
 \end{align} where $H$ is the number of components, each of which represents a \emph{group}, $\Xi = \{\xi_1, 
 \ldots, \xi_H\}$, $\xi_h = \{\mathscr{m}_h, \mathscr{C}_h\}$ and $\sum_{h = 1}^H \tau_h = 1$. In particular, \begin{align*}
     \text{N}_h(\mathscr{y}_b \mid \xi_h) = \prod_{i \in \mathscr{I}_b} \text{N}(y_i \mid \xi_h).
 \end{align*}
 
 Placing commonly used conjugate priors on $\Xi$ and $\tau$ \begin{align}
(\tau_1, \ldots, \tau_{H} ) &\sim \text{Dir}(\alpha_0, \ldots, \alpha_0), \label{equ:prior3}\\
\mathscr{m}_h \mid \mathscr{C}_h \sim \text{N}(\mathbf{0}, \mathscr{C}_h)\quad \text{and}\quad
\mathscr{C}_h &\sim \text{IW}\left(\nu_0, \mathbf{S}_0\right) \quad h = 1, \ldots, H \label{prior3},
\end{align} 
yields a Bayesian model. The refinement procedure is not sensitive to the specific choice of prior parameters because each item contains many observations.

The computation based on \eqref{equ:group_mod} involves two key ideas. The first one, which is the use of a \emph{reference subset}, is a simple idea to ensure that the mapping from updated subcomponent labels to cluster labels results in reasonably contiguous clusters. Specifically, we randomly select one subset as the reference, denoted by $\mathscr{r}^*$ and let the number of mixture components $H = \mathscr{L}_{\mathscr{r}^*}$ be the number of items on the reference. The group allocation $\mathscr{z}_{\cdot}$ of an item is simply the index in the mixture model given by \eqref{equ:group_mod} to which it has been assigned.  With each item in $\mathscr{r}^*$ representing a different mixture component in \eqref{equ:group_mod}, we are effectively aligning each item from across all subsets (including the reference) to an item in $\mathscr{r}^*.$ The updated item label $\tilde{\mathscr{z}}_{\cdot}$ is the index of the item to which it is aligned in the reference. In other words, the updated item label for any item matched to $(\mathscr{r}^*, h)$ is $\tilde{\mathscr{z}}_{\cdot} = h$. Once the item has been aligned with an item in the reference subset, the associated cluster label is updated to the cluster label of this item in the reference, $\ceil*{\frac{\tilde{\mathscr{z}}_{\cdot}}{L}}.$ For example, if three subcomponents (i.e. $L = 3$) are used to approximate a cluster and the item index set in the reference subset $\mathscr{r}^*$ is $\{(\mathscr{r}^*, 1), (\mathscr{r}^*,2), (\mathscr{r}^*, 5),( \mathscr{r}^*, 9)\}$ based on the first indexing rule, then a possible mapping between the items and the mixture components is given in Table \ref{tab:item_component_map}. If an item $b$ from a different subset is assigned to group 3 $(\mathscr{z}_b = 3)$, it is equivalently aligned with $(\mathscr{r}^{*}, 5)$; then its updated label is the index of this item in $\mathscr{r}^{*},$ or $\tilde{\mathscr{z}}_b = 5$. Consequently its associated cluster label is updated to $2 \,( = \ceil*{\frac{5}{3}})$. If item $(\mathscr{r}^{*}, 5)$ is assigned to group 2 (i.e. $\mathscr{z}_{(r^*, 5)} = 2$), then it is aligned with $(\mathscr{r}^{*}, 2)$ and its associated cluster label is updated to 1, which amounts to the merge of cluster 1 and 2 in the reference subset. The final clustering estimate is not sensitive to the choice of reference, since we refine many local clustering samples, with each provided a separate reference drawn at random. 
%Then for item $b$, its updated cluster label is given by $\ceil*{\frac{\tilde{\mathscr{z}}_b}{L}}.$
%Then we can map the updated subcomponent labels
%$\mathscr{z}_b$ to the cluster level via $\ceil*{\frac{\mathscr{z}_b}{L}}.$
\begin{table}[]
\centering
\caption{An example dictionary of item indices, mixture components and cluster labels \label{tab:item_component_map}}
\begin{tabular}{|c|cccc|}
\hline
Item index & $(\mathscr{r}^*, 1)$ & $(\mathscr{r}^*,2)$ & $(\mathscr{r}^*, 5)$ & $( \mathscr{r}^*, 9)$ \\ \hline
Mixture component in \eqref{equ:group_mod} (group label) & 1 & 2 & 3 & 4 \\ \hline
Cluster label & 1 & 1 & 2 & 3\\\hline
\end{tabular}
\end{table}

%In addition, once each subcomponent has be matched to a subcomponent on the reference node, it can be placed in the cluster associated with that subcomponent. This will solve the label ambiguity problem across clusters and generate a unified clustering for all data points.

%sampling  $\mathscr{r}^*$ at random so that its item allocations can be used as the reference for group assignment to ensure consistent labeling across subsets. Specifically, each item in subset $\mathscr{r}^*$ is mapped to a different component in \eqref{equ:group_mod}, with $H = \mathscr{L}_{\mathscr{r}^*}.

The second key idea is the use of a \emph{collapsed Gibbs sampler}. Our model specification, \eqref{equ:group_mod}, \eqref{equ:prior3} and \eqref{prior3}, enables derivation of a collapsed Gibbs sampler, where we integrate out the model parameters  $\{\Xi,\tau\}$ from the joint posterior and only update the latent group allocation through MCMC sampling; such a sampler in general accelerates convergence to the posterior distribution. This sampler, as we shall illustrate, can be implemented in parallel and depends on the data only through three summary statistics of each item.

%The result is not sensitive to the particular choice of the reference subset, because we refine many samples and each refinement starts with a separate random draw of the reference.
 %Mapping the group label back to the cluster level results in alignment of cluster labels and automatic merging of cluster 3 and 5 in subset 2; see Figure \ref{fig:refine_illus2} for an illustration of an example.
 
%Finally, we can use a randomly selected reference node to create consistent labels across the subsets. Specifically, a node $r^* \in \{1, \ldots, R\}$ is randomly drawn so that $\mathbf{s}_{r^*}$ serves as the reference to which we jointly align the group allocations $z_b$ $(b = 1, \ldots, B)$ and then the cluster allocations associated with item $b$ are updated to be $\ceil*{\frac{\mathscr{z}_b}{L}}$, a consistent set of labels used across workers. 
%ased on their degree of overlap, one natural model employ with $H$ components, each of which represents a

To facilitate explanation of the sampler, we introduce the following notation. Let the vector of group allocations and the vector of updated item labels be $\mathbf{z}$ and $\tilde{\mathbf{z}}$ respectively. Using the second item indexing rule, for example, $\mathbf{z} = \{\mathscr{z}_1, \mathscr{z}_2, \ldots, \mathscr{z}_B\}.$
Denote the associated quantities of group $h$, including the data, sample size, first and second moment, by $\mathscr{Y}_h$, $\mathscr{N}_{h}$, $\bar{\mathscr{Y}}_{h}$ and $\mathscr{S}_{h}$ respectively. Following the second indexing rule, $\mathscr{Y}_h = \{\mathscr{y}_b: \mathscr{z}_b =  h, b = 1, \ldots, B\}.$ Subscript ${h \setminus b}$ represents group $h$ without taking item $b$ into account. Let $\mathscr{Q}_{b, h} = (\mathscr{N}_{h \setminus{b}}, \bar{\mathscr{Y}}_{h \setminus{b}}$ $\mathscr{S}_{h \setminus{b}}).$

%We first (randomly) assign each item in the reference to a different component/group in \eqref{equ:group_mod}. 
The key quantity in updating the group allocation is the posterior probability of item $b$ being assigned to group $h$: 
\begin{align}P(\mathscr{z}_b = h \mid \mathbf{\mathbf{z}}_{\setminus{b}}, \mathcal{Y}) \propto P(\mathscr{z}_{b} = h \mid \mathbf{z}_{\setminus{b}}) p(\mathscr{y}_b \mid \mathscr{Y}_{h \setminus{b}}), \quad h = 1, \ldots, H; b = 1, \ldots, B \label{equ:prob},\end{align}
where the first term is
\begin{align}
P(\mathscr{z}_{b} = h \mid \mathbf{z}_{\setminus{b}}) = \frac{\Gamma(N + H\alpha_0 - \mathscr{n}_b )\Gamma(\mathscr{N}_{h \setminus{b}} + \mathscr{n}_b + \alpha_0) }{\Gamma(N + H\alpha_0) \Gamma(\mathscr{N}_{h \setminus{b}} + \alpha_0)},\quad b = 1, \ldots, B,\label{equ:assignprob}
\end{align}
and the second term is the joint marginal density of observations in item $b$, which is given by a product of $t$-densities:
\begin{align}
p(\mathscr{y}_b \mid \mathscr{Y}_{h \setminus{b}}) = \prod_{j \in \mathscr{I}_b}  \text{t}_d\left({y}_{j}\,\Bigg|\, \mathbf{m}_{h \setminus{b}},  \frac{\kappa_{h \setminus{b}} + 1}{\kappa_{h \setminus{b}}(\nu_{h \setminus{b}}- d + 1)}\mathbf{S}_{h \setminus{b}}, \nu_{h \setminus{b}} - d + 1\right),\quad b = 1, \ldots, B, \label{equ:jointt}
\end{align}
where \begin{align*}\kappa_{h \setminus{b}} &= 1 + \mathscr{N}_{h \setminus{b}}, \quad \nu_{h \setminus{b}} = \nu_0 + \mathscr{N}_{h \setminus{b}}, \\ \mathbf{m}_{h \setminus{b}} &= \frac{\mathscr{N}_{h \setminus{b}} \bar{\mathscr{Y}}_{h \setminus{b}}}{\kappa_{h \setminus{b}}} \quad\text{and}\quad \mathbf{S}_{h \setminus{b}} = \mathbf{S}_0 + \mathscr{N}_{h \setminus{b}}\mathscr{S}_{h \setminus{b}} - \kappa_{h \setminus{b}} \mathbf{m}_{h \setminus{b}} \mathbf{m}_{h \setminus{b}}^T. \end{align*}
The details of the derivation are included in Appendix \ref{appendix:collapsed}.

Computation of \eqref{equ:assignprob} and \eqref{equ:jointt} requires statistics $\mathscr{N}_{h \setminus{b}}$, $\bar{\mathscr{Y}}_{h \setminus{b}}$ $\mathscr{S}_{h \setminus{b}}$, h = 1, \ldots, H, which are simply functions of sample size $\mathscr{n}_b$, mean $\bar{\mathscr{y}}_b$ and second moment ${\mathscr{S}}_b$, $b = 1, \ldots, B$, since
\begin{align}
\mathscr{N}_{h \setminus{b}} &=  \left(\sum_{i \in b \cup \{j: \mathscr{z}_j = h\}}{\mathscr{n}_i}\right) - \mathscr{n}_b,\label{equ:groupsize}\\
\bar{\mathscr{Y}}_{h \setminus{b}}&= \left[ \left(\sum_{i \in b \cup \{j: \mathscr{z}_j = h\}}{\mathscr{n}_i} \bar{\mathscr{y}}_i\right) - \mathscr{n}_b\bar{\mathscr{y}}_b\right]\Bigg/\mathscr{N}_{h \setminus{b}},\label{equ:groupmean}\\
\mathscr{S}_{h \setminus{b}} &= \left[ \left(\sum_{i \in b \cup \{j: \mathscr{z}_j = h\}}{\mathscr{n}_i} {\mathscr{S}}_i\right) - \mathscr{n}_b{\mathscr{S}}_b\right]\Bigg/\mathscr{N}_{h \setminus{b}}.\label{equ:group2mom}
\end{align}
Therefore, it suffices to have the workers communicate these sufficient statistics, sample size $\mathscr{n}_b$, mean $\bar{\mathscr{y}}_b$ and second moment ${\mathscr{S}}_b$ ($b = 1, \ldots, B$) to the master. Then the master evaluates $\{\mathscr{Q}_{b, h}: \forall b \text{ and } h\}$ and their function \eqref{equ:assignprob}. $\{\mathscr{Q}_{b, h}: \forall b \text{ and } h\}$  are then communicated to the appropriate worker $\mathscr{r}_b = \{r: \mathscr{y}_b \subset \mathcal{Y}_r, r = 1,\ldots, R\}$ to evaluate \eqref{equ:jointt} for all $b$, which can be completed in an embarrassingly parallel manner. The resulting values will again be communicated to the master for evaluation of \eqref{equ:prob} and updating latent group allocations $\mathbf{z},$ based on which the associated cluster labels of subset $r$ are updated to be $\tilde{\mathbf{c}}_r := \{\tilde{c}_{r1}, \ldots, \tilde{c}_{rn_r}\}$, with \begin{align}
   \tilde{c}_{ri} = \ceil*{\frac{\tilde{\mathscr{z}}_{(r, j)}}{L}}, \text{ for }  i \in \mathscr{I}_{(r,j)}\label{equ:updated_clust}
\end{align} 
where $i = 1, \ldots, n_r$ and $r = 1, \ldots, R.$ The resulting $\tilde{\mathbf{c}}$ is considered an approximate sample from $p(\mathbf{c} \mid \mathcal{Y})$. See Algorithm \ref{alg:align} for a clear outline and the flow chart below for an intuitive illustration of the above steps. We find refining $100$ local clustering samples $\mathbf{c} = \{\mathbf{c}_1$, \ldots, $\mathbf{c}_R\}$ by running {one} iteration of Algorithm \ref{alg:align} is sufficient for excellent performance. 

\begin{enumerate}
    \item{A sample of subcomponent allocations} 
    \begin{center}
    \includegraphics[width = 0.8\textwidth]{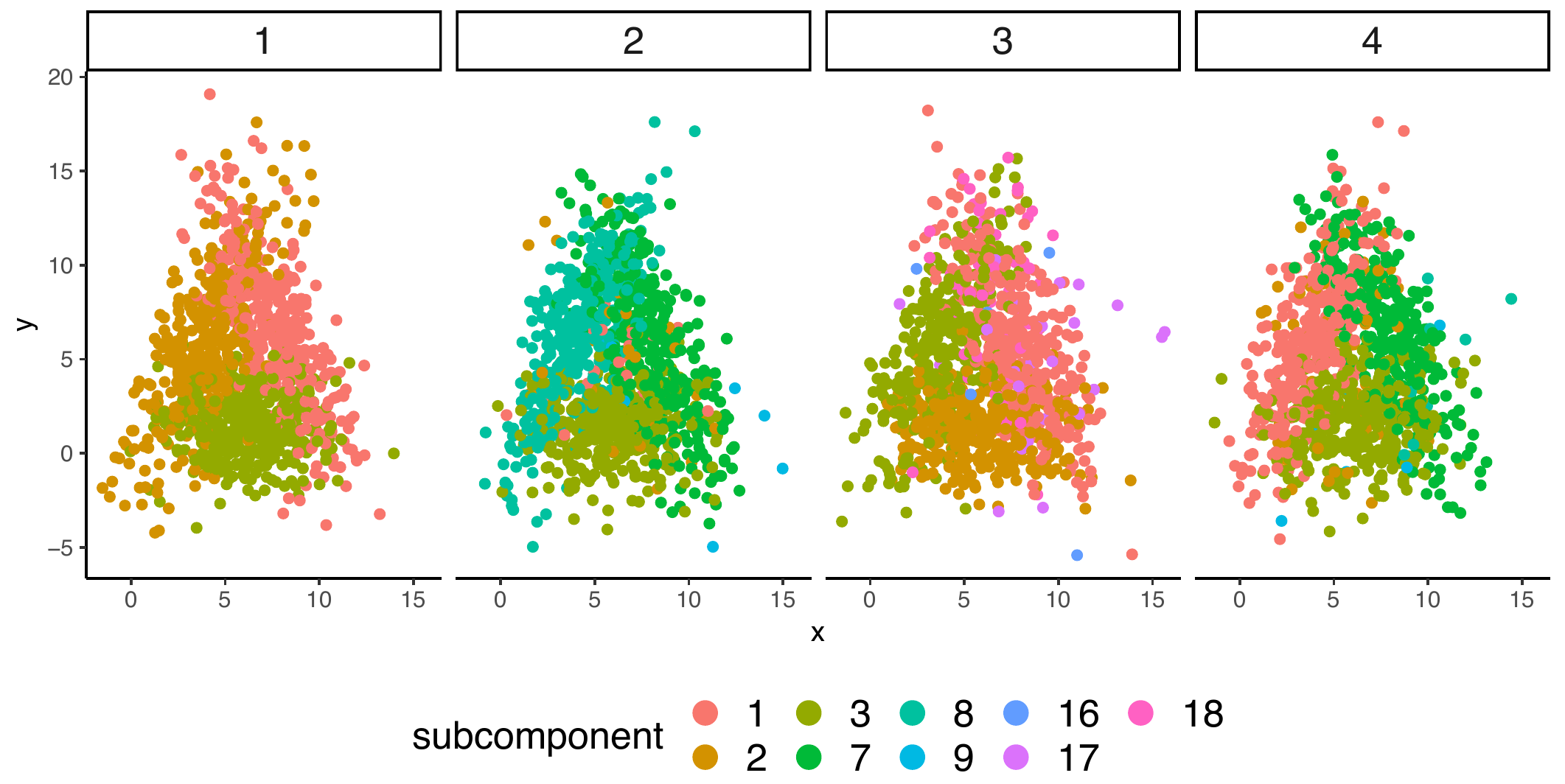}
    \end{center}
    \item{Random choice of the reference: $\mathscr{r}^{*} = 3$}
    \begin{center}
\begin{minipage}{\textwidth}
  \stackunder{\begin{minipage}[t]{0.63\textwidth}
    \centering
    \includegraphics[width = 0.4\textwidth]{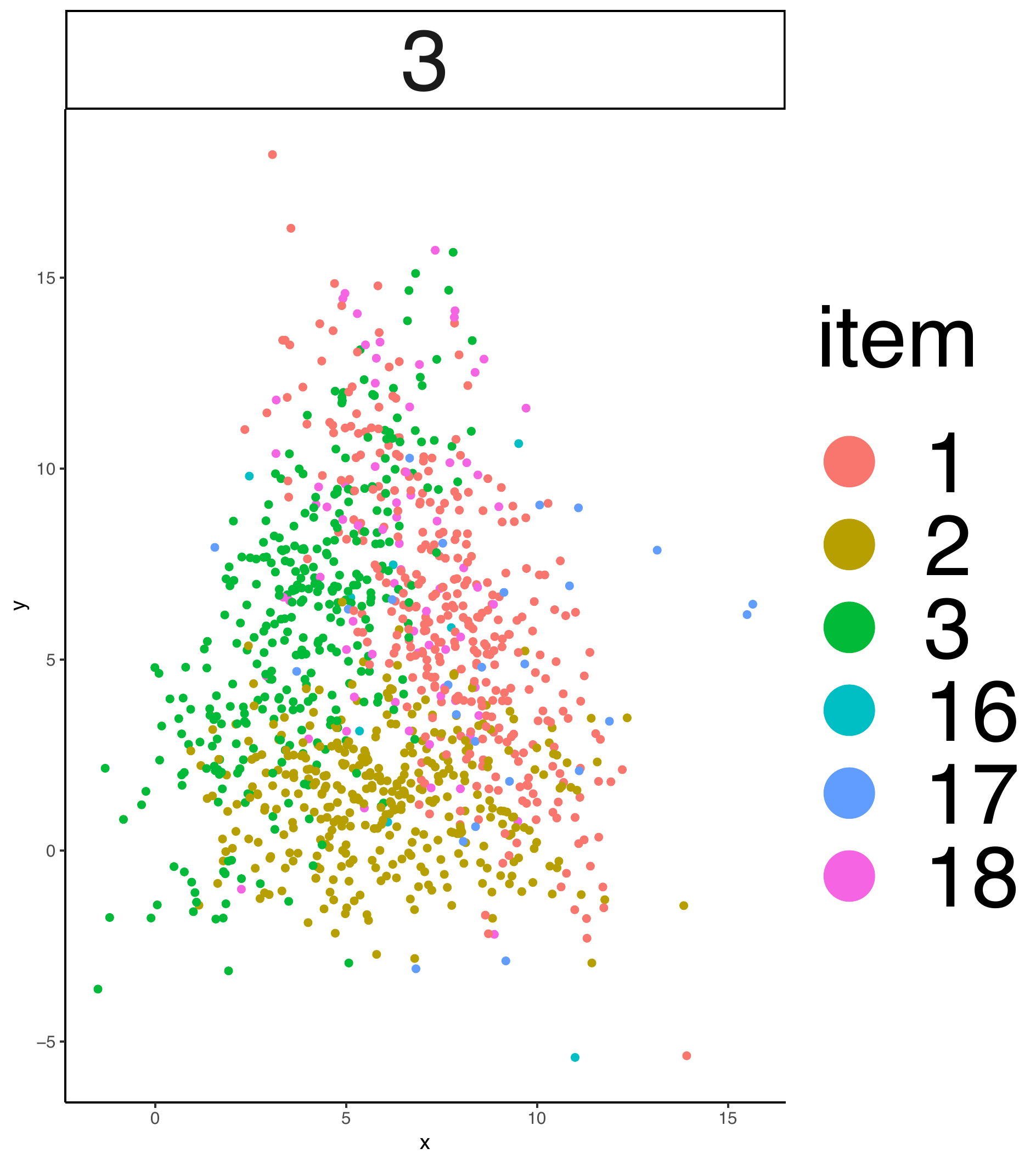}
      \end{minipage}}
      % dont need this caption
    {\begin{minipage}[t]{0.46\textwidth}
    \captionof{figure}{The sample of subcomponent allocations in the reference subset}
  \end{minipage}}
  \hfill
  \stackunder{\begin{minipage}[t]{0.37\textwidth}
    %\mytable
    \vspace{-4cm}
    \centering
   \begin{tabular}{l|l}
Item index & Group \\ \hline
(3,18)     & 1     \\
(3,17)     & 2     \\
(3,16)     & 3     \\
(3,3)      & 4     \\
(3,2)      & 5     \\
(3,1)      & 6    
  \end{tabular}
    \end{minipage}}
    {\begin{minipage}[t]{0.37\textwidth}
      \captionof{table}{A bijective map between the item indices and groups\label{tab:refine_map}}
    \end{minipage}}
\end{minipage}
\end{center}

     \item{Initialization to the closest item in the reference}
     \begin{center}
        \includegraphics[width = 0.8\textwidth]{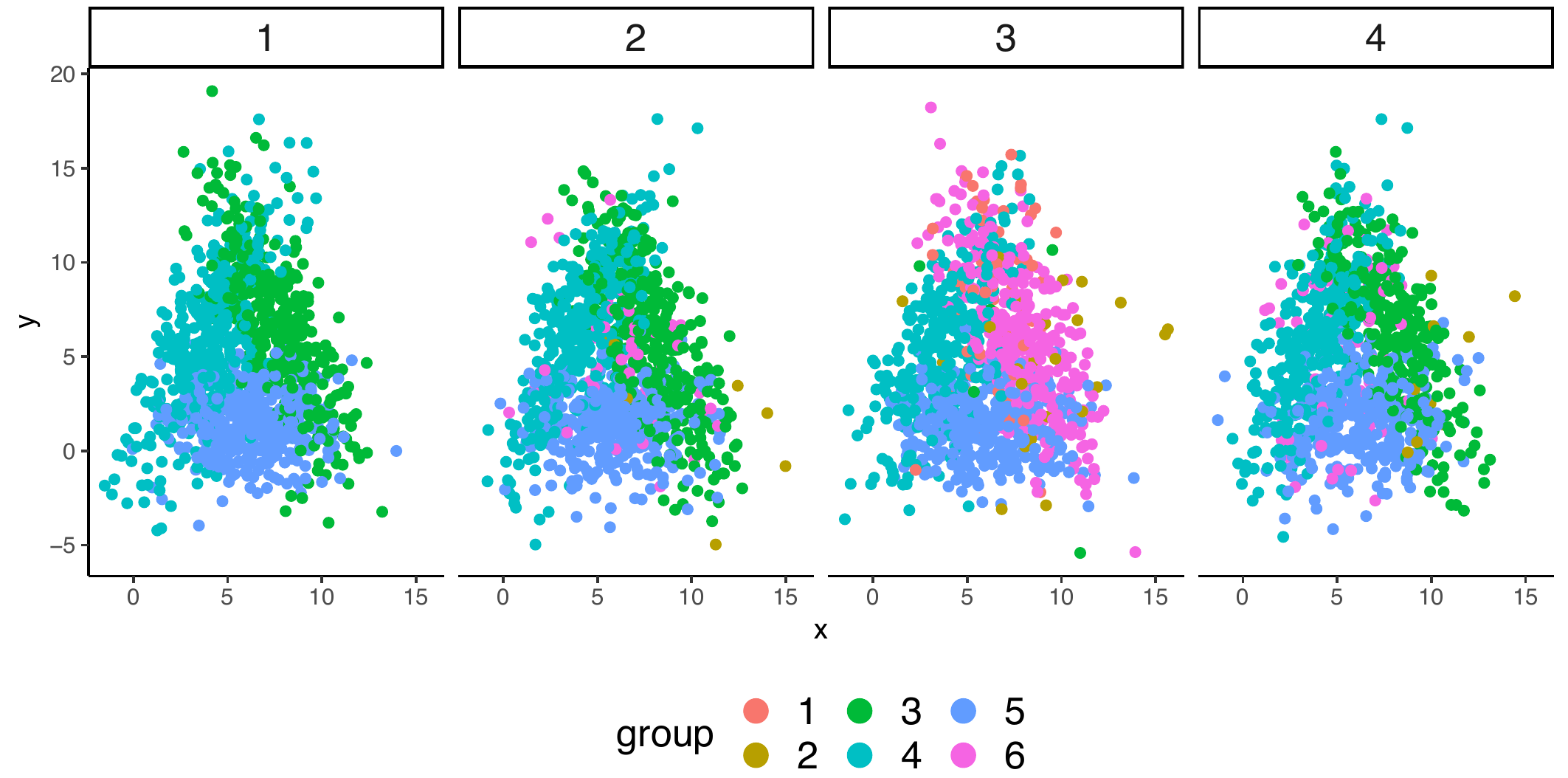}
     \end{center}
          \item{Results from running one iteration of collapsed Gibbs sampler}
          \begin{center}
          \includegraphics[width = 0.8\textwidth]{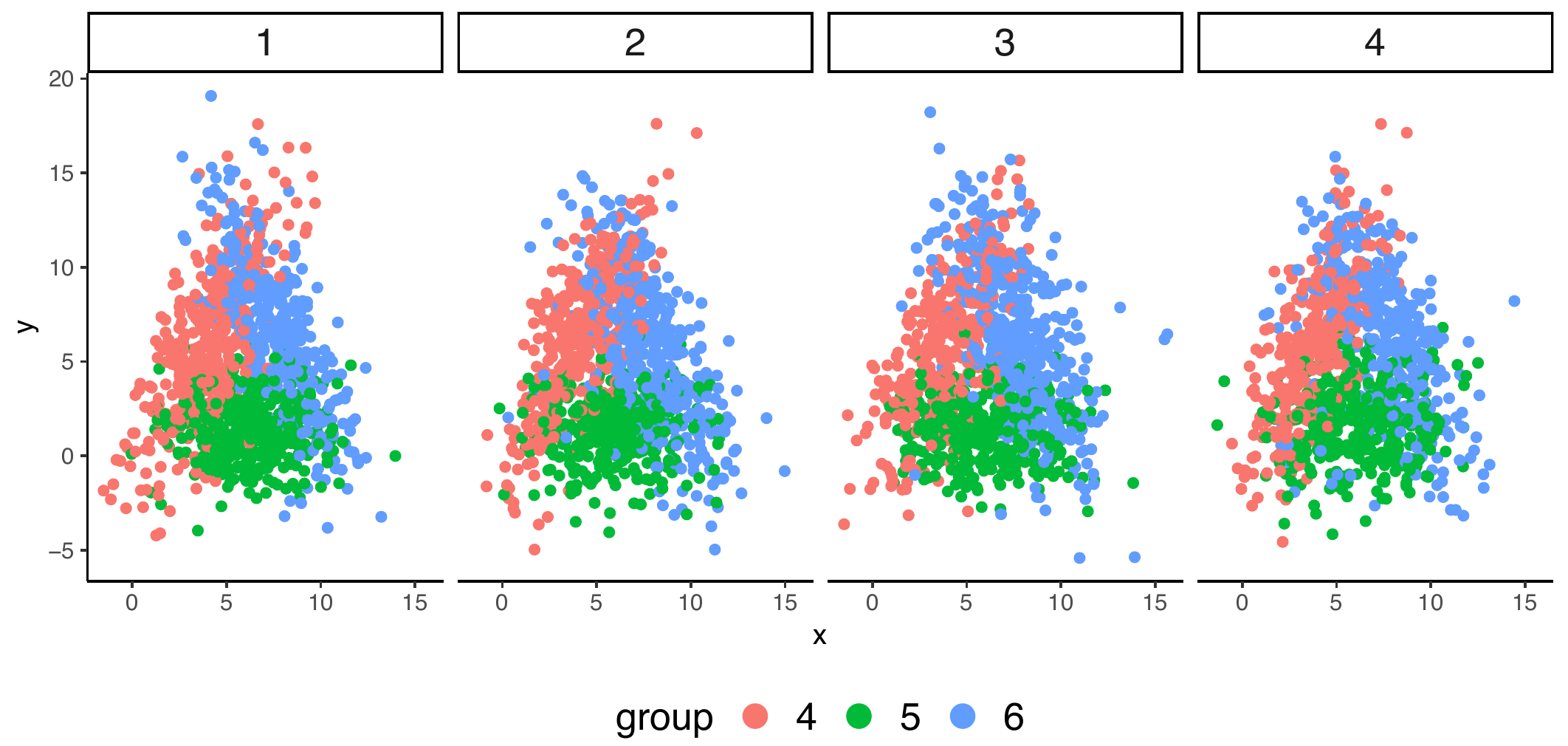}
          \end{center}
           \item{Item labels updated based on Table \ref{tab:refine_map}}
           \begin{center}
               \includegraphics[width = 0.8\textwidth]{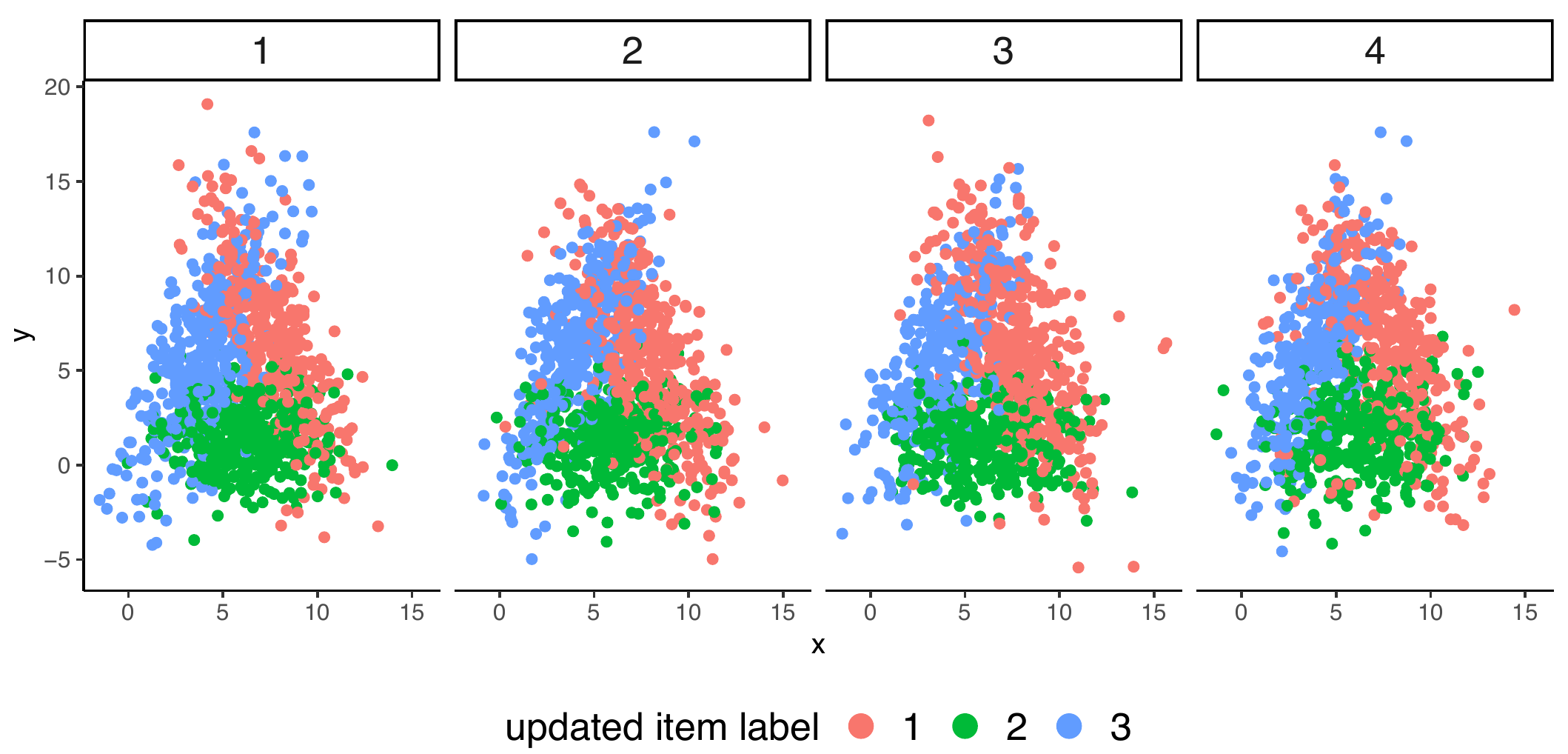}
           \end{center}
              \item{Cluster labels updated based on \eqref{equ:updated_clust}}
              \begin{center}
              \includegraphics[width = 0.8\textwidth]{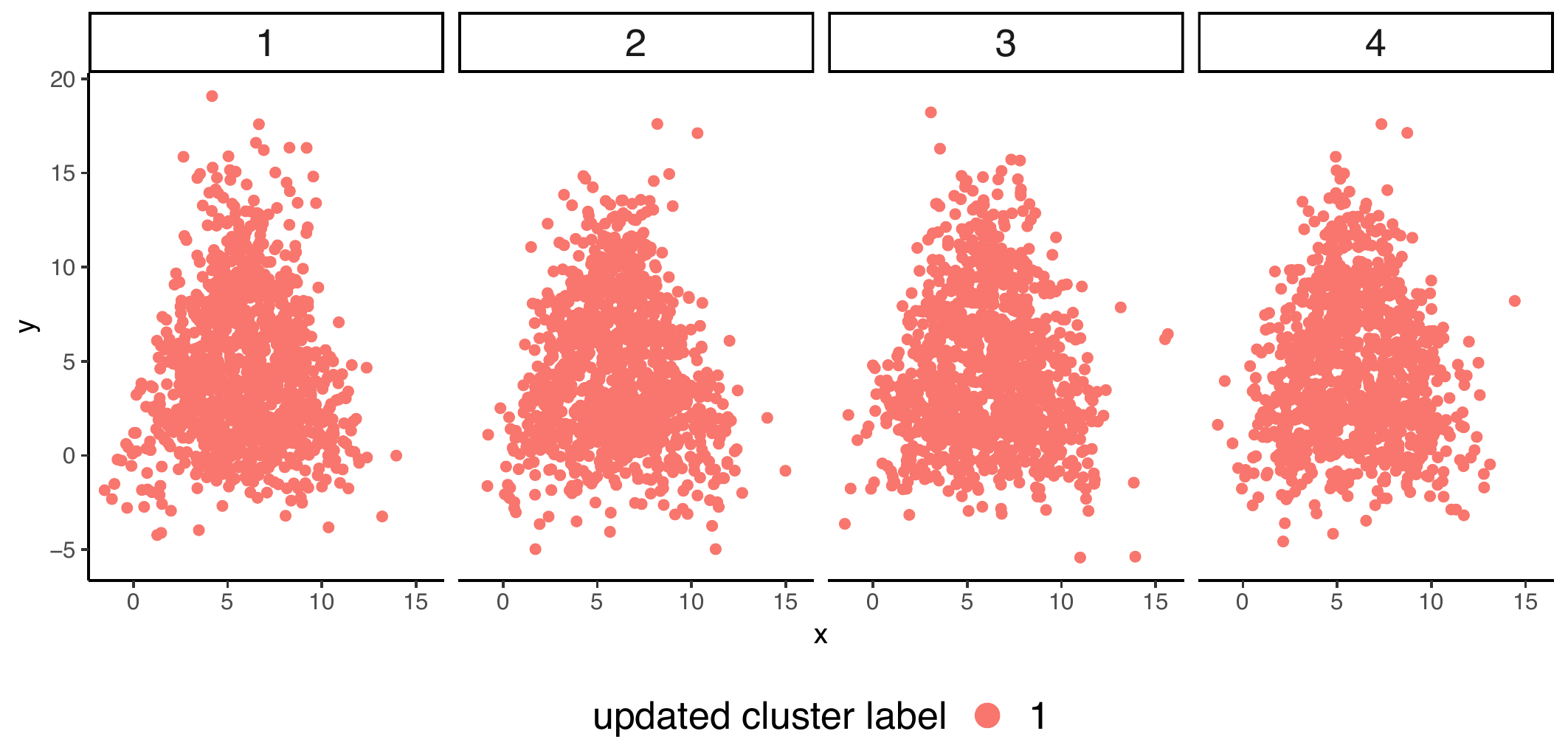}
              \end{center}
\end{enumerate}

%	%	\State (Parallel Collapsed Gibbs sampler for an overfitted Gaussian mixture model)
		%\Statex \Comment { \%comment: servers[] con	tains the index of servers whose         data rate are sorted in descending order\%}
\begin{algorithm}
	\caption{Global Cluster Refinement (for one iteration of local cluster samples)}
	% \newline(Parallel Collapsed Gibbs sampler for a Gaussian mixture model)
	\label{alg:align}
	\hspace*{\algorithmicindent} \textbf{Input:} Summary statistics $\{(\mathscr{n}_b, \bar{\mathscr{y}}_b, \mathscr{S}_b):  b = 1, \ldots, B\}$ sent from Workers to Master \\
 \hspace*{\algorithmicindent} \textbf{Output:} Updated cluster labels $\tilde{\mathbf{c}} =\{\tilde{\mathbf{c}}_1, \ldots, \tilde{\mathbf{c}}_R\}$
	\begin{algorithmic}[1]
%	\Procedure{Euclid}{$a,b$} \Comment{The g.c.d. of a and b}
		%\State \textbf{From workers to master:} item size $\mathscr{n}_b$, mean $\bar{\mathscr{y}}_b$ and second moment $\mathscr{S}_b$, $b = 1, \ldots, B$.
		\State \textbf{On Master:}
		\State Draw a subset to be the reference $\mathscr{r}^*$ at random
		\State $H \gets \mathscr{L}_{\mathscr{r}^*}$
		\State Determine a bijective map $\mathscr{g}: \text{items in subset } \mathscr{r}^* \to \{1, \ldots, H\}$
	%	\State $\mathscr{z}_j \gets \mathscr{g}(j), \forall \text{ item }j$ in subset $\mathscr{r}^*$\Comment{Assign each item in $\mathscr{r}^*$ to a different component}
		\For{$b = 1$ to $B$}
	\LineComment{Initialize $\mathscr{z}_b$ by identifying its closest item in the reference subset}
		\State $\mathscr{z}_b \gets \mathscr{g}\left({\arg\min_j}\Vert \bar{\mathscr{y}}_b - \bar{\mathscr{y}}_{(\mathscr{r}^*, j)}\Vert\right)$ 
		\EndFor
		%\State Initialize $\mathbf{z} = \{\mathscr{z}_1,\ldots,$ %\mathscr{z}_B\}$ to by mapping each item to the subcomponent in the reference node whose mean is closest in the Mahalanob
		\For{$b = 1$ to $B$ }
		    \For{$h = 1$ to $H$}
		        \State  Compute $\mathscr{Q}_{b,h}$ using \eqref{equ:groupsize}, \eqref{equ:groupmean} and \eqref{equ:group2mom}
		        \State 	Send $\mathscr{Q}_{b, h}$ to Worker $\mathscr{r}_b = \{r: \mathscr{y}_b \subset \mathcal{Y}_r, r = 1,\ldots, R\}$
		        \State Compute $P(\mathscr{z}_b = h \mid \mathbf{z}_{\setminus{b}})$ using \eqref{equ:assignprob}
		    \EndFor 
		\EndFor 
		%	\For{$1$ iteration \do}
		%	\For {$b=1$ to $B$ \do}
		%	\State Remove ${\mathfrak{y}}_b$'s statistics from component $\tilde{s}_b$. \Comment{Old assignment for $\mathcal{\tilde{Y}}^b$}
		%	\State \textbf{From the center node} $N_{h \setminus{b}}$, $\bar{\mathscr{Y}}_{h \setminus{b}}$ $\mathscr{S}_{h \setminus{b}}$ \textbf{to the worker node} where item $b$ belongs, for all $h$ and $b$.
		\State \textbf{On Workers:}
		\ParFor{Worker $r= 1$ to $R$}
		\For{all item $(r, j)$}
		\For{$h = 1$ to $H$}
		\State Compute $p(\mathscr{y}_{(r,j)} \mid \mathscr{Y}_{h\setminus (r, j)})$ using $\mathscr{Q}_{(r,j),h}$ and \eqref{equ:jointt}
		\State Send $p(\mathscr{y}_{(r,j)} \mid \mathscr{Y}_{h\setminus (r,j)})$ to Master
		\EndFor 
		\EndFor 
		\EndParFor
		\State \textbf{On Master:}
		\For{$b = 1$ to $B$}
		\For{$h = 1$ to $H$}
		\State  $P(\mathscr{z}_b = h \mid \mathbf{z}_{\setminus{b}}, \mathcal{Y}) \gets p(\mathscr{y}_b \mid \mathscr{Y}_{h\setminus b})P(\mathscr{z}_b = h \mid \mathbf{z}_{\setminus{b}})$
		\EndFor
		\State Draw $\mathscr{z}_b$ from a categorical distribution with probability vector \\$\mathbf{p} \propto \left[P(\mathscr{z}_b = 1\mid \mathbf{z}_{\setminus{b}}, \mathcal{Y}), \ldots, P(\mathscr{z}_b = H\mid \mathbf{z}_{\setminus{b}}, \mathcal{Y})\right]$
		\State $\tilde{\mathscr{z}}_b \gets \mathscr{g}^{-1}(\mathscr{z}_b)$
		\State Send $\tilde{\mathscr{z}}_b$ to Worker $\mathscr{r}_b = \{r: \mathscr{y}_b \subset \mathcal{Y}_r, r = 1,\ldots, R\}$
		\EndFor
			\State \textbf{On Workers:}
			\ParFor{Worker $r= 1$ to $R$}
		\For{all item $(r, j)$}
		%	\State $\tilde{\mathscr{c}}_{(r, j)}  \gets \ceil*{\frac{\tilde{\mathscr{z}}_{(r,j)}}{L}}$
			 \State $\tilde{c}_{ri} \gets \ceil*{\frac{\tilde{\mathscr{z}}_{(r, j)}}{L}}, \text{ for } i \in \mathscr{I}_{(r,j)}$
		\EndFor
		\State\Return{$\tilde{\mathbf{c}}_r$}
		\EndParFor
		%		\Statex\Comment{ \%comment: find the equation corresponding the serverIndex        from the mapping at the File Server\%} 
		%		\State        $eqn= equation(serverIndex[i])$ 
		%		\Statex\Comment{ \%comment: try insert equation into Z using OFG\%} 
	%	\EndProcedure
	\end{algorithmic}
\end{algorithm}
%Finally, we can use a randomly selected reference node to create consistent labels across the subsets. Specifically, a node $r^* \in \{1, \ldots, R\}$ is randomly drawn so that $\mathbf{s}_{r^*}$ serves as the reference to which we jointly align the group allocations $z_b$ $(b = 1, \ldots, B)$ and then the cluster allocations associated with item $b$ are updated to be $\ceil*{\frac{\mathscr{z}_b}{L}}$, a consistent set of labels used across workers. 

Based on Algorithm \ref{alg:align}, we see that items originating from the same cluster can possibly be assigned to different groups and end up forming separate clusters, and items originating from different clusters can potentially be assigned to the same or adjacent groups, which become subcomponents of the same cluster. This shows that our global cluster refinement algorithm is flexible enough to allow both cluster merging and splitting.

\subsection{Global Clustering Estimation}
Decision theory-based point estimates of clustering $\mathbf{c}^*$ provide an elegant solution to the problem of producing a single clustering from a posterior \citep{WadGha17}. The optimal clustering estimate is one that minimizes the posterior expectation of a loss function:
\begin{align}
\mathbf{c}^* = \underset{\hat{\mathbf{c}}}{\arg \min}\, \mathbb{E}[L(\mathbf{c}, \hat{\mathbf{c}})\mid\mathcal{Y}].\label{equ:risk}
\end{align}
$\mathbb{E}[L(\mathbf{c}, \hat{\mathbf{c}})\mid\mathcal{Y}]$ is often simplified to be a function that depends on the posterior \emph{only} through the posterior similarity matrix, which is defined to be an $N$ by $N$ matrix with entry $(i,j)$ being the posterior probability that data points $i$ and $j$ are assigned to the same cluster $P(c_i = c_j \mid \mathcal{Y}).$ Since $P(c_i = c_j \mid \mathcal{Y})$ can be readily estimated by the proportion of posterior samples that cluster data points $i$ and $j$ together, it has been a common practice to estimate the posterior similarity matrix in order to obtain $\mathbf{c}^*$. Such a approach, however, becomes computationally infeasible for big data of size $N$: the $N$ by $N$ posterior similarity matrix incurs both large storage and computational costs, with the latter being $O(N^2)$ for each iteration involved in creating the posterior similarity matrix.

We propose a simple solution that exploits an inherent property of any loss function defined on the space of partitions and as a result is general enough to accommodate any eligible loss. Since clustering is invariant to the permutation of data point indices, any loss $L(\mathbf{c}, \hat{\mathbf{c}})$ must be a function of the \emph{joint counts} $N_{ij} = |{C}_i \bigcap \widehat{C}_j|$, which is the number of data points in both $C_i$, the set of data indices in cluster $i$ under $\mathbf{c}$, and $\widehat{C}_j$, the set of data indices in cluster $j$ under $\hat{\mathbf{c}}$, $i = 1, \ldots, \text{number of clusters in } \mathbf{c}$ and $j = 1, \ldots, \text{number of clusters in } \hat{\mathbf{c}}$  \citep{binder:1978:cluster}. These joint counts can be easily obtained through parallel computation.
%The fact that clustering is invariant to the permutation of data point indices implies that a loss function  ; these joint counts can be easily obtained through parallel computation. Another advantage of this method is that it can accommodate any loss function defined on the space of partitions, due to its direct use of the definition.
%For example, under Binder's loss $\mathbf{B}(\mathbf{c}, \hat{\mathbf{c}}) = \sum_{n < n^\prime} \mathbbm{1}(c_n = c_{n^\prime})\mathbbm{1}(\hat{c}_n \neq \hat{c}_{n^\prime}) + \mathbbm{1}(c_n \neq c_{n^\prime})\mathbbm{1}(\hat{c}_n = \hat{c}_{n^\prime})$, 
%\begin{align*}\mathbb{E} [\text{B}(\mathbf{c}, \hat{\mathbf{c}})\mid \mathcal{Y}] =  \sum_{n < n^\prime} \left(\mathbbm{1}(\hat{c}_n = \hat{c}_{n^\prime}) - P\left(c_n = c_{n^\prime}\mid \mathcal{Y}\right)\right)^2,
%\end{align*}where $
%P\left(c_n = c_{n^\prime}|\mathcal{Y}\right)$ can be readily estimated by the proportion of posterior samples in which $n$ and $n^\prime$ are allocated to the same cluster. Therefore, a common practice to estimate $\mathbf{c}^*$ is through the posterior similarity matrix. 
%in the set $\mathscr{T}$, where $\mathscr{T}$ contains all the iterations involved in creating the posterior similarity matrix.

Without loss of generality, we demonstrate our algorithm based on the variation of information (VI) loss. The VI loss is an information theoretic criterion for comparing two clustering structures and is defined as: \begin{align}
\text{VI}(\mathbf{c}, \hat{\mathbf{c}}) &= \text{H}(\mathbf{c})+ \text{H}(\hat{\mathbf{c}}) - 2 \text{I}(\mathbf{c},\hat{\mathbf{c}}) \nonumber \\
&= \sum_{i = 1}^{k_N}\frac{N_{i+}}{N}\log\left(\frac{N_{i+}}{N}\right) + \sum_{j = 1}^ {\hat{k}_N} \frac{N_{+j}}{N}\log\left(\frac{N_{+j}}{N}\right) - 2\sum_{i = 1}^{k_N}\sum_{i = 1}^{\hat{k}_N} \frac{N_{ij}}{N} \log\left(\frac{N_{ij}}{N}\right),\label{equ:videf} 
\end{align}
where $H$ is the entropy function and $I$ is the mutual information \citep{meila:2007:viloss}, $k_N$ and $\hat{k}_N$ are the number of clusters in $\mathbf{c}$ and $\hat{\mathbf{c}}$ respectively, and $N_{i+}$ and $N_{+j}$ represent the counts in $C_i$ under $\mathbf{c}$ and $C_j$ under $\hat{\mathbf{c}}$ respectively. The posterior expected VI loss can be simplified to:
\begin{align}
\mathbb{E} [\text{VI}(\mathbf{c}, \hat{\mathbf{c}})\mid\mathcal{Y}] = \sum_{n = 1}^N \log\left(\sum_{n = 1}^N \mathbbm{1}(\hat{c}_{n^\prime} = \hat{c}_n) \right) - 2 \sum_{n = 1}^N \mathbb{E}\left(\log(\sum_{n^\prime = 1}^N\mathbbm{1}(c_{n^\prime} = c_n, \hat{c}_{n^\prime} = \hat{c}_n))\mid\mathcal{Y}\right) \label{equ:vi1} 
%\\
\end{align} up to a constant. To improve the computational efficiency, \citet{WadGha17} developed a lower bound of \eqref{equ:vi1} and reduced the computation cost to $O(N^2)$ for a given clustering candidate $\hat{\mathbf{c}},$ which still remains astronomical for our case. 
%Because estimating \eqref{equ:vi1} is computationally intensive---$O(\vert \mathscr{T} \vert N^2)$ for a given clustering candidate $\hat{\mathbf{c}}$, where $\mathscr{T}$ is the set of all iterations involved in creating the posterior similarity matrix---\citet{WadGha17} used Jensen's inequality to upper bound \eqref{equ:vi1} by $\sum_{n = 1}^N \log \left(\sum_{n^\prime = 1}^N \mathbbm{1}(\hat{c}_{n^\prime} =  \hat{c}_{n}) \right)- 2 \sum_{n = 1}^N \log\left(\sum_{n^\prime = 1}^NP(c_{n^\prime} = c_{n} | \mathcal{Y})\mathbbm{1}(\hat{c}_{n^\prime} = \hat{c}_n)\right)$ and had computational cost reduced to $O(N^2)$; this still remains astronomical for our case.

Alternatively, let $\mathscr{T} \in \{1, \ldots, T\}$ be the set of iterations adjusted in the global cluster refinement step. Consider a random subset $\mathscr{M} \subset \mathscr{T}$ and clustering candidates $\hat{\mathbf{c}} \in \{\tilde{\mathbf{c}}^{(t)}, t \in  \mathscr{M}\}$. The optimal point estimate of clustering is given by: \begin{align*}
    \mathbf{c}^* = \underset{\tilde{\mathbf{c}}^{(t)}, t \in \mathscr{M}}{\arg\min \:}  \widehat{\mathbb{E} [\text{VI}(\mathbf{c}, \tilde{\mathbf{c}}^{(t)})\mid\mathcal{Y}] }.
\end{align*}
Specifically, the posterior expected VI loss given clustering candidate $\hat{\mathbf{c}}$ can be estimated according to the definition \eqref{equ:videf} by:
\begin{align}
\widehat{\mathbb{E} [\text{VI}(\mathbf{c}, \hat{\mathbf{c}})\mid\mathcal{Y}] }= \sum_{j = 1}^ {\hat{k}_N} \frac{N_{+j}}{N}\log\left(\frac{N_{+j}}{N}\right) - \frac{2}{\vert \mathscr{T} \vert }\sum_{t \in \mathscr{T}}\sum_{i = 1}^{k_N^{(t)}}\sum_{j = 1}^{\hat{k}_N} \frac{N_{ij}^{(t)}}{N} \log\left(\frac{N_{ij}^{(t)}}{N}\right),\label{equ:viloss_est}
\end{align}
where superscript $(t)$ represents iteration $t$. Note that both $N_{ij}^{(t)}$ and $N_{+j}$ can be computed in parallel, because the refined cluster allocations $\{\tilde{\mathbf{c}}^{(t)}, t \in \mathscr{T}\}$ make the following relationships hold:
\begin{align*} N_{ij}^{(t)} = \sum_{r = 1}^R N_{r, ij}^{(t)}, \quad N_{+j} = \sum_{r = 1}^R N_{r, +j},  \end{align*}
where $N_{ij}^{(t)}$ and $N_{+j}$ are computed on the master using the statistics $N_{\cdot, ij}^{(t)}$ and $N_{\cdot, +j}$ communicated from the workers. See Algorithm \ref{alg:cluster_estimate} for a detailed outline of the above procedures.

For each clustering candidate, the evaluation of $\{N_{ij}^{(t)}: \forall t \in \mathscr{T}, i = 1, \ldots, k_N^{(t)}, j = 1, \ldots, \hat{k}_N\}$ and  $\{N_{r, +j}, j = 1, \ldots, \hat{k}_N\}$ is $O(n\vert\mathscr{T}\vert)$, where $n := \max_{r} n_r$ is the largest sample size on any worker. The cost of communicating these statistics to the master depends on the total number of variables being transferred, which is $\hat{k}_N R\left(\sum_{t \in \mathscr{T}}k_N^{(t)}  + 1\right)$ for every clustering candidate. Although $\hat{k}_N R \sum_{t \in \mathscr{T}}k_N^{(t)} $ is approximately linear in the number of iterations, $\vert \mathscr{T}\vert$ can be chosen to be much smaller than the total number of iterations for sampling local clustering. Our experiments show superior clustering performance with $\vert \mathscr{T} \vert  = 100$ and $\vert \mathscr{M} \vert = 20$.

% (CHANGE) 
%$O(\vert\mathscr{M}\vert k_N^{(t^\prime)} + \vert\mathscr{M}\vert k_N^{(t^\prime)}\vert \mathscr{T}\vert k_N^{(t)})$
%The clustering candidates $\hat{\mathbf{c}}$ are a random subset of $\{1, \ldots, T\}$,
%For iteration $t$, worker $r$ first evaluates $N_{r, ij}^{(t)}$ and $N_{r, +j}$ (for $i = 1, \ldots, k_N^{(t)}$ and $j = 1, \ldots, \hat{k}_N$) in parallel, $r = 1, \ldots, R$; such evaluation incurs computational cost $O(nT)$, where $n := \max_{r} n_r$ the largest sample size on any worker. The workers then transmit the statistics to the master; for a clustering candidate with $\hat{k}_N$ clusters, the amount of data communication is $\vert \mathscr{T} \vert \hat{k}_N k_N^{(t)}R$, where $k_N^{(t)} \leq K$. Next, the master node evaluates all $N_{ij}^{(t)}$ and $N_{+j}$ based on  \eqref{equ:counts_relationship} and then the quantity in \eqref{equ:viloss_est}.  

\begin{algorithm}
	\caption{Global Clustering Estimation}
	% \newline(Parallel Collapsed Gibbs sampler for a Gaussian mixture model)
	\label{alg:cluster_estimate}
	\hspace*{\algorithmicindent} \textbf{Input:} Output of refined samples of local cluster allocations from Algorithm \ref{alg:align}\\
 \hspace*{\algorithmicindent} \textbf{Output:} Optimal clustering estimate $\mathbf{c}^*$
	\begin{algorithmic}[1]
%	\Procedure{Euclid}{$a,b$} \Comment{The g.c.d. of a and b}
		%\State \textbf{From workers to master:} item size $\mathscr{n}_b$, mean $\bar{\mathscr{y}}_b$ and second moment $\mathscr{S}_b$, $b = 1, \ldots, B$.
				\State \textbf{On Workers:}
						\ParFor{Worker $r = 1$ to $R$}
						\LineComment{Iterate through the set of clustering candidates}
		\For{iteration $t^\prime \in \mathscr{M}$} 
		\For{cluster $j = 1$ to $k_N^{(t^\prime)}$} \Comment{$k_N^{(t^\prime)}$: number of clusters in candidate $t^\prime$}
		\State Compute $N_{r, +j}^{(t^\prime)}$
		\State Send $N_{r, +j}^{(t^\prime)}$ to Master
		\LineComment{Iterate through the set of refined local cluster allocations samples}
		\For{iteration $t \in \mathscr{T}$}
		\For{cluster $i = 1,\ldots, k_{N}^{(t)}$}
		\State Compute $N_{r, ij}^{(t, t^\prime)}$
		\State Send $N_{r, ij}^{(t, t^\prime)}$ to master
		\EndFor
		\EndFor
		\EndFor
		\EndFor
		\EndParFor
	\State\textbf{On Master:}
	\For{iteration $t^\prime \in \mathscr{M}$} 
	\State $\text{Term}_1^{(t^\prime)} \gets 0$
	\Comment{$\text{Term}_1^{(t^\prime)}:$ the first term in \eqref{equ:viloss_est}}
	\State $\text{Term}_2^{(t^\prime)} \gets 0$\Comment{$\text{Term}_2^{(t^\prime)}:$ the second term in \eqref{equ:viloss_est}}
		\For{cluster $j = 1$ to $k_N^{(t^\prime)}$}
			\State $N_{+j}^{(t^\prime)} \gets \sum_{r = 1}^R N_{r, +j}^{(t^\prime)}$
			\State $\text{Term}_1^{(t^\prime)} \gets \text{Term}_1^{(t^\prime)} +\left({N_{+j}^{(t^\prime)}}/{N}\right)\log\left({N_{+j}^{(t^\prime)}}\Bigg/{N}\right)$ 
	\For{iteration $t \in \mathscr{T}$} 
		\For{cluster $i = 1,\ldots, k_{N}^{(t)}$}
	\State $N_{ij}^{(t, t^\prime)} \gets \sum_{r = 1}^R N_{r, ij}^{(t, t^\prime)}$
	\State $\text{Term}_2^{(t^\prime)} \gets \text{Term}_2^{(t^\prime)} + \left({N_{ij}^{(t, t^\prime)}}/{N}\right) \log\left({N_{ij}^{(t, t^\prime)}}\Bigg/{N}\right)$
	\EndFor
	\EndFor
	\EndFor
	\State	$\widehat{\mathbb{E}\left[\text{VI}(\mathbf{c}, \tilde{\mathbf{c}}^{(t^\prime)} \mid \mathcal{Y}\right]} \gets \text{Term}_1^{(t^\prime)} - \text{Term}_2^{(t^\prime)}$
	\EndFor
	\State $\mathbf{c}^* \gets\underset{\tilde{\mathbf{c}}^{(t^\prime)}, t^\prime \in \mathscr{M}}{\arg\min \:}  \widehat{\mathbb{E} [\text{VI}(\mathbf{c}, \tilde{\mathbf{c}}^{(t^\prime)})\mid\mathcal{Y}] }$
	\State \Return $\mathbf{c}^*$
	\end{algorithmic}
\end{algorithm}
\subsection{Sampling Model Parameters}
This section only applies if one is interested in quickly classifying future subjects, density estimation or generating from posterior predictive for inference. These goals make drawing (approximate) posterior samples of model parameters $(\Theta, \eta)$ necessary. For example, to classify a new subject, the \emph{Bayes classifier}---evaluating the posterior probability of belonging to each cluster and assigning to the cluster that yields the highest posterior probability---is commonly used and depends on the model parameters.

%$j$: $P\left(\mathbf{y} \in \text{cluster } j\right) = \frac{{\eta}_j f_{j}(\mathbf{y} | {\theta}_j)}{\sum_{k = 1}^{{k_N}} {\eta}_k f_k(\mathbf{y} | {\theta}_k)}, \label{equ:BayClass}$ $j = 1, \ldots, {k}_N$  
%subcomponent-specific parameters (i.e. $\omega_{kl}$, $\mu_{kl}$ and $\Sigma_{kl}$) and other 
The general idea of our algorithm is to sample the model parameters conditional on the cluster and subcomponent assignments associated with the optimal clustering estimate found in the last step (i.e. global clustering estimation), allowing the MCMC updates to depend on the data only through summary statistics. These statistics include subcomponent sizes $\vert i: \tilde{s}_i = l, \tilde{c}_i = k \vert$, the sum of squares $\sum_{i: \tilde{s}_i = l, \tilde{c}_i = k} y_i y_i^{T}$ and the data sums $\sum_{i: \tilde{s}_i = l, \tilde{c}_i = k}y_i$, $i = 1, \ldots, N.$ Refer to Appendix \ref{appendix:mcmc} for the details of MCMC updates.
% $l =1, \ldots, L$, $k = 1, \ldots, k_N$ 

Such an approach has several advantages. First, subcomponent assignments are aligned across all subsets through global cluster refinement, which makes the following relationships hold: 
\begin{align*}
\sum_{i: y_i \in \mathcal{Y}}\mathbbm{1}_{\tilde{s}_i = l, \tilde{c}_i = k} &= \sum_{r = 1}^R \sum_{i:y_i \in \mathcal{Y}_r} \mathbbm{1}_{\tilde{s}_i = l, \tilde{c}_i = k} ,\\
\sum_{\substack{y_i \in \mathcal{Y} \\\tilde{s}_i = l, \tilde{c}_i = k}} y_i y_i^{T} &= \sum_{r = 1}^R \sum_{\substack{y_i \in \mathcal{Y}_r, \\\tilde{s}_i = l, \tilde{c}_i = k}}  y_i y_i^{T}, \\ \sum_{\substack{y_i \in \mathcal{Y}, \\\tilde{s}_i = l, \tilde{c}_i = k}} y_i &=\sum_{r = 1}^R \sum_{\substack{y_i \in \mathcal{Y}_r, \\\tilde{s}_i = l, \tilde{c}_i = k}} y_i, \quad\\ l &=1, \ldots, L, k = 1, \ldots, k_N^*, \,\text{where } k_N^* \text{ is the number of clusters in }\mathbf{c}^*.\nonumber 
\end{align*}
These relationships justify our parallel algorithm: the workers first compute the relevant statistics in parallel and then communicate them to the master node for summing. Such communication only occurs once; because conditional on the subcomponent assignments, the summary statistics are fixed throughout the MCMC updates. Second, the fixed subcomponent assignments and the use of summary statistics significantly reduce computation per iteration and lead to faster convergence due to reduced dependence between iterations. Third, posterior estimates of parameters are often non-identifiable in finite mixture models due to label switching, a phenomenon which occurs when exchangeable priors are placed on the parameters. It is particularly challenging to resolve, despite many available post-sampling relabeling schemes \citep{stephens2000, SperrinJW10, PapIli10, Pap14, RodWalker14}, for overfitted mixture models, as superfluous clusters may merge or overlap with other ones, or be empty. By fixing the latent cluster assignments, we bypass label switching at the cluster level, eliminating the need to post process MCMC outputs. See Appendix \ref{appendix:alg_sampling} for an outline of the algorithm.

\section{Experiments \label{sec:experiment}}
In this section, we provide extensive numerical experiments to assess the performance of DIB-C, as measured by the scalability of the computation time and classification performance with the data size and number of computing workers. We use training sets of size $N = 12$ thousand, $120$ thousand, and $1$ million with test sets of $3$ thousand, $30$ thousand, and $250$ thousand observations, respectively ($20\%$ of the total data size). All the experiments are conducted on the Duke Compute Cluster; due to limitations of the computing cluster, we only benchmark the time of DIB-C up to 30 workers and performance up to 120 workers.

When the data are distributed to a single node, global cluster refinement is not necessary; that is, a full MCMC for sampling (local) clustering is immediately followed by (global) cluster estimation. The prior setup is shared by all experiments from the same data set. In the sampling local clustering step, we run 1000 iterations of MCMC across all experiments, with the first 500 being burn-ins. Then we refine 100 iterations of local clustering samples, from which we randomly select 20 to be the clustering candidates considered for the global clustering estimation based on the variation of information loss. To sample the model parameters, we run 2000 iterations of MCMC conditional on the optimal subcomponent and cluster assignments.
\subsection{Experimental setup}
\subsubsection{Synthetic data sets}

\begin{figure}
	\centering
	\includegraphics[width = 0.4 \textwidth]{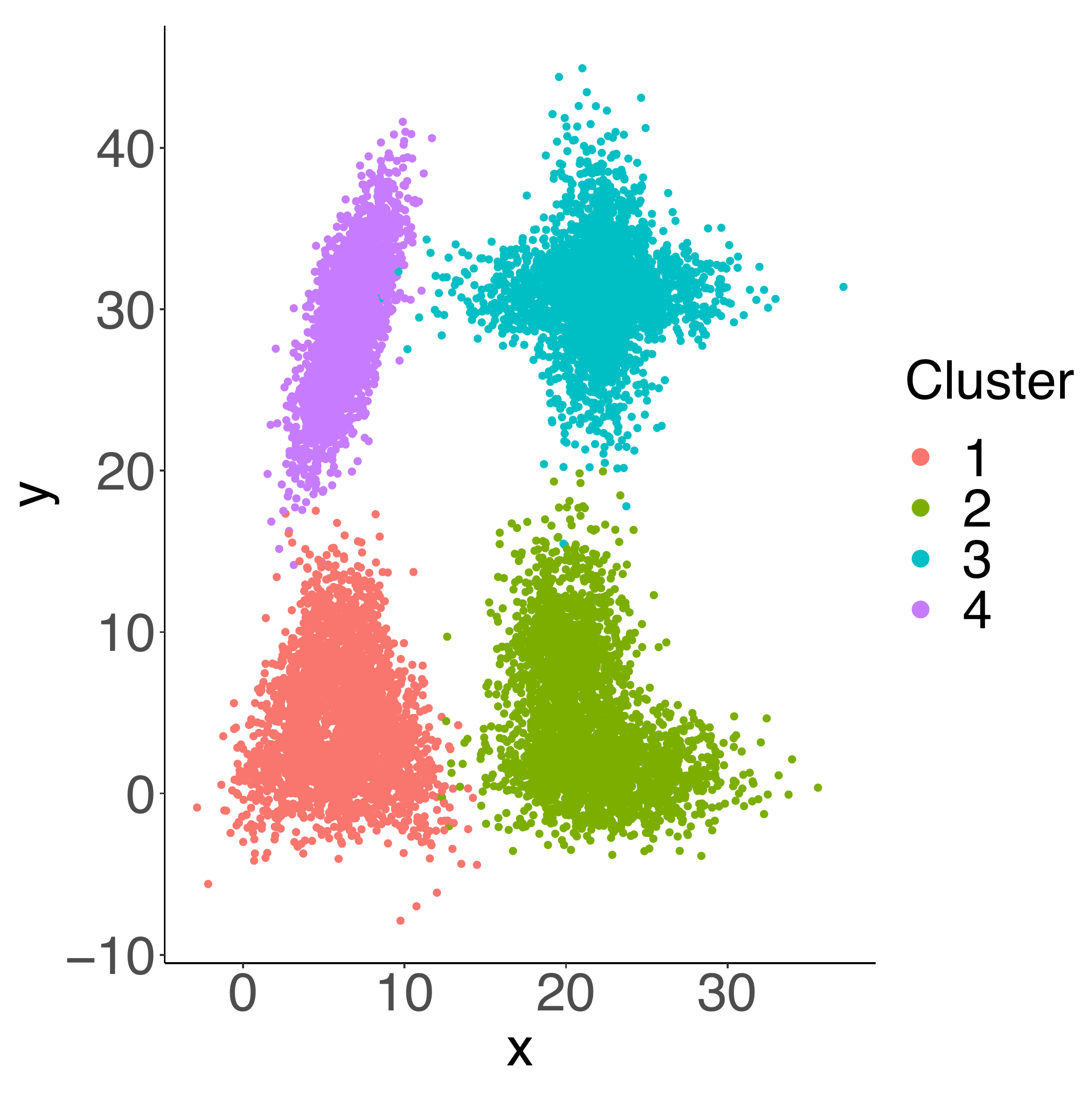}
	\includegraphics[width = 0.45\textwidth]{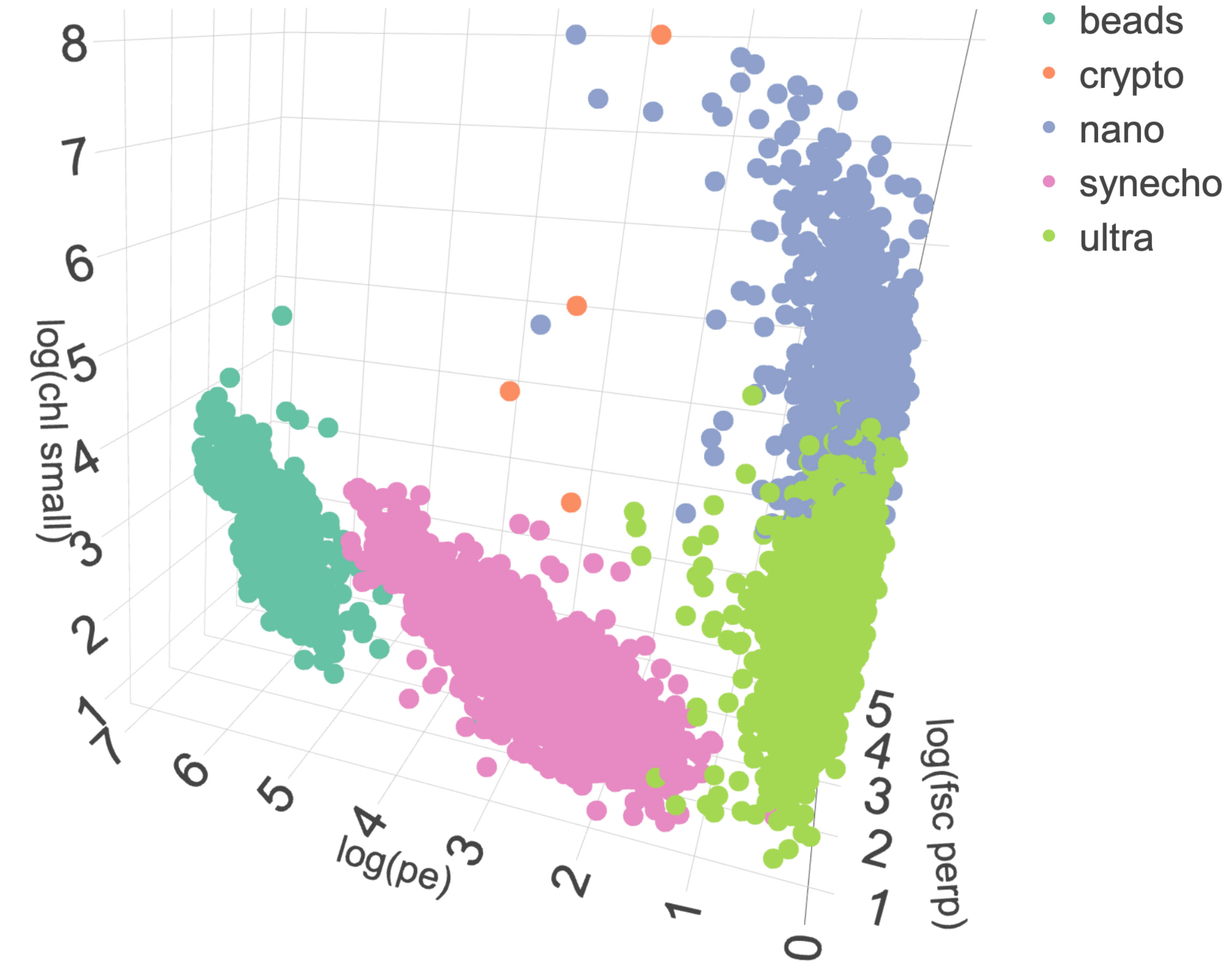}
	\caption{Left: the scatter plot of a synthetic data set simulated with $N = 12,000$ observations. Right: the scatter plot of 10,000 observations randomly drawn from the 1 million training set of the flow cytometry data, where chl small, pe and fsc perp represent chlorophyll level, phycoerythrin level, and forward scatter respectively \label{fig:scatterplot}}
\end{figure}

The simulation setup originated in \cite{baudryetal2010} and was later augmented by \cite{Wallietal17}. The data sets contain four clusters of varying shapes, including one triangle, one L, one cross, and one ellipse. They are generated from an eight-component Gaussian mixture in $\mathbb{R}^2$ with component means	\begin{align*}\left(\mu_1,\ldots, \mu_8\right) = 
\begin{pmatrix}
6&4& 8& 22.5 &20& 22& 22& 6.5\\
1.5& 6& 6& 1.5& 8& 31& 31& 29
	\end{pmatrix},
\end{align*}covariance matrices
\begin{align*}
 \scalemath{0.95}{
\begin{matrix}
\Sigma_1 =\begin{pmatrix}4.84&0\\0&2.89\end{pmatrix},  &\Sigma_2 = \begin{pmatrix}3.61&5.05\\5.05&14.44\end{pmatrix}, 
&\Sigma_3 = \begin{pmatrix}3.61&-5.05\\-5.05&14.44\end{pmatrix},  &
\Sigma_4 = \begin{pmatrix}12.25&0\\0&3.24\end{pmatrix},\\ \Sigma_5 = \begin{pmatrix}3.24&0\\0&12.25\end{pmatrix}, &\Sigma_6 = \begin{pmatrix}14.44&0\\0&2.25\end{pmatrix} ,
&\Sigma_7 = \begin{pmatrix}2.25&0\\0&17.64\end{pmatrix},
&\Sigma_8 = \begin{pmatrix}2.25&4.20\\4.20&16.00 \end{pmatrix},
\end{matrix}}
\end{align*} cluster weight vector $\eta = \left(\frac{1}{4}, \frac{1}{4}, \frac{1}{4}, \frac{1}{4}\right)$ and subcomponent weight vectors $\omega_{1} = \left(\frac{1}{3}, \frac{1}{3},\frac{1}{3}\right)$, $\omega_2  = \left(\frac{1}{2}, \frac{1}{2}\right)$, $\omega_3  = \left(\frac{1}{2}, \frac{1}{2}\right)$, $\omega_4 = 1$. Figure \ref{fig:scatterplot} shows a scatter plot of a data set simulated from this setup. In our experiment, we simulate 10 data sets for each sample size. To estimate the model, we let $K = 10$ and $L = 3$.
\subsubsection{Flow cytometry data}
%These quantities are benchmarked against the reference obtained from full MCMC with \emph{true} cluster labels supplied.
We consider high-frequency, continuous flow cytometry data collected from particles in aquatic environments by SeaFlow instruments \citep{hyrkas2016}. Specifically, the SeaFlow instruments continuously sample surface seawater, generating a time series of cytometry samples (one every 3 minutes) containing measurements of the optical properties, including light scatter and intrinsic fluorescence, of small phytoplankton cells during a research cruise \citep{Jeremy15clusterflowcytometry}. The data set \citep{seaflowdata} contains four optical measurements: forward scatter, side scatter, phycoerythrin level, and chlorophyll level. One important problem of interest is to classify phytoplankton cells based on their optical measurements.
%fsc_perp: perpendicular scatter of a particle (forward scatter)?
%pe: phycoerythrin measurement of a particle
%chl_small: chlorophyll measurement of a particle

%Each phytoplankton cell was evaluated for its light scatter and intrinsic fluorescence. Light scatter relevant measurements are forward scatter and side scatter, which are proportional to the cell size and cell complexity respectively; fluorescence relevant measurements include red fluorescence (characteristic of chlorophyll) and orange fluorescence (characteristic of phycoerythrin) \citep{Sosik2010}. 

Currently the dominant classification method in this application area is manual gating, where a scientist manually identifies the physical boundaries for clusters of cells. This process is subjective and can be infeasible given the massive amount of data collected during a research cruise \citep{hyrkas2016}. In addition, properties of the same phytoplankton species may vary as environmental conditions (e.g. daylight) change over time and space \citep{Sosik2010}, leading to variation in the shapes and centers of the clusters.

Figure \ref{fig:scatterplot} shows a scatter plot of the three variables (side scatter, phycoerythrin and chlorophyll) that explain most variation in the data. The plot includes a random sample from the first one million observations with the variables on a logarithmic scale. We note that synecho, a species of phytoplankton, has visually disjoint components, but nano and ultra, two distinct species, are not well separated. In addition, crypto is a rare cluster with wide spread. All these factors increase the difficulty of clustering this data set.

A recent state-of-the-art clustering method for flow cytometry data first segments the data based on visual inspection of a change point in cluster formation and independently clusters these segments using a Gaussian mixture model with a pre-determined number of clusters \citep{hyrkas2016}. In evaluating the clustering performance, they excluded observations whose assignment probability to any cluster is less than 0.7 and benchmarked the remaining observations against their manually-gated labels; this procedure resulted in an average F-measure of 0.864. Our method, in contrast, does not require a pre-determined number of clusters as an input or exclude observations based on their assignment probabilities. 

To avoid time dependent variation in the clusters and given limited computing resources, we only use the first $1.25$ million observations from the data set and similarly benchmark against the manually-gated labels. Since the data are partially labelled, the training set contains around $3.5\%$ unlabeled data for each data size setting $N$. These unlabeled data are used in model fitting, but are excluded in clustering performance evaluation. Noticing that each variable is \emph{highly} right skewed, we transform the four variables to their logarithmic scales for ease of approximation. We set the upper bound of the number of clusters $K = 8$ and $L = 2$ for the experiments.
\subsection{Results}
This section includes experiment results for both the synthetic and flow cytometry data. Reported results include benefits from the global clustering refinement algorithm, classification performance, simulations from the posterior predictive distributions, and computation times.

The classification performance is evaluated for both the training and test sets by three frequently used cluster validation metrics---accuracy, adjusted Rand index (ARI) and F-measure; see Appendix \ref{appendix:metricdef} for the definitions. When two partitions agree perfectly, accuracy, ARI and F-measure take value 1; in general, higher values of these statistics represent better clustering performance. To evaluate the test set performance, we first find an optimal mapping that minimizes the number of mismatches between the labels of the optimal clustering estimate of the training set and the manually-gated labels, so that each label refers to the same cluster in both the training and test set. The extra clusters identified in the estimates are coded as \emph{unknown1}, \emph{unknown2} and so on. All results, unless explicitly stated, are averaged across 10 replicated experiments. 
%The posterior summaries include the posterior mean and $95\%$ credible intervals associated with the cluster weight $\eta_k$, center $\mu_k =  \sum_{l} \omega_{kl} \mu_{kl}$ and spread $\Sigma_k := \sum_{l = 1}^L \omega_{kl}\Sigma_{kl} + \sum_{l = 1}^L \omega_{kl}\mu_{kl}\mu_{kl}^T - \mu_k \mu_k^T$. 
\subsubsection{Benefits from global cluster refinement algorithm}
\begin{figure}
\centering
	\includegraphics[width =0.7\textwidth]{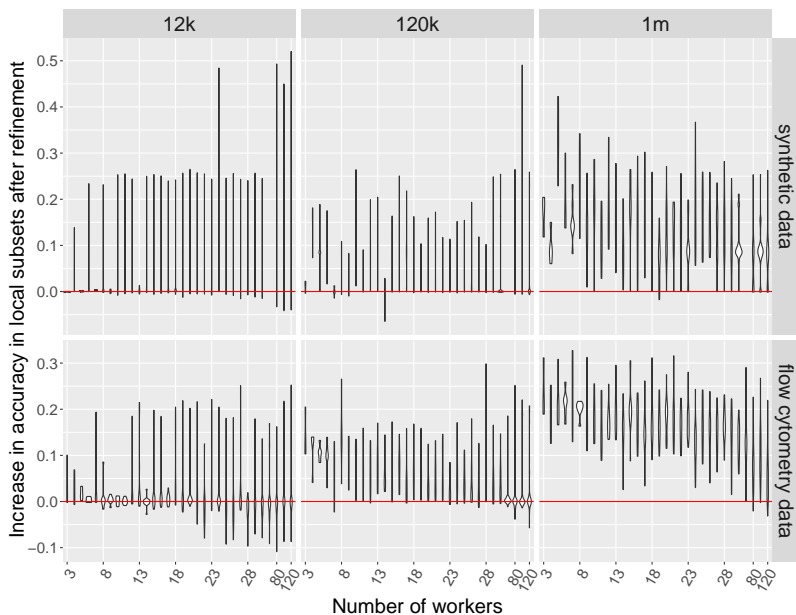}
	\caption{Increase in classification accuracy after refinement in local subsets of the training sets from the synthetic (top row) and the flow cytometry data (bottom row). 12k, 120k and 1m indicate the data size of the training sets. For each data and node setting, we randomly select one of ten replicates and include a violin plot of $R$ statistics, where $R$ is the number workers (or subsets). \label{fig:syndat_accuracydelta}}
\end{figure}
\begin{figure}
	\centering
\includegraphics[width=0.49\textwidth]{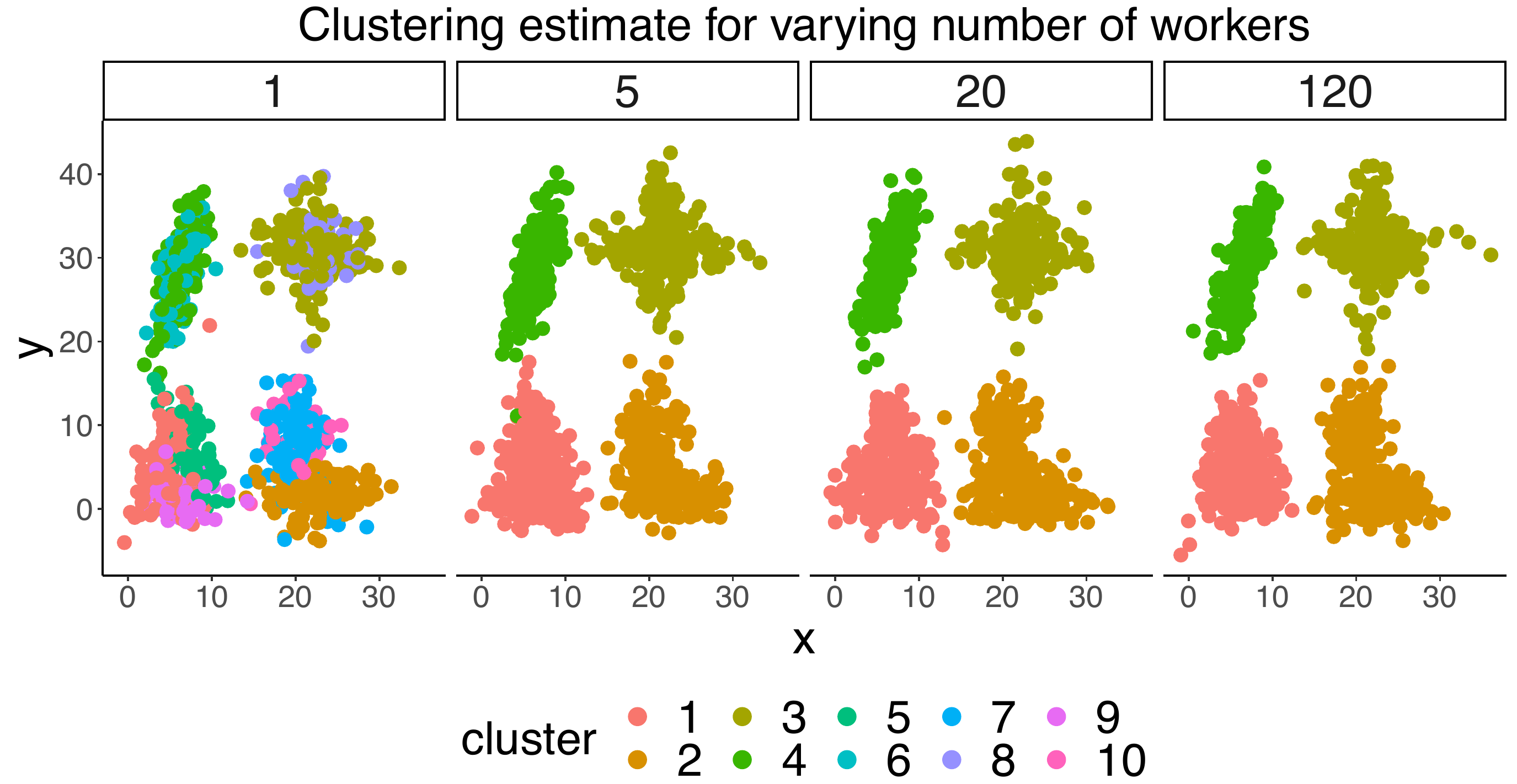}
\includegraphics[width=0.49\textwidth]{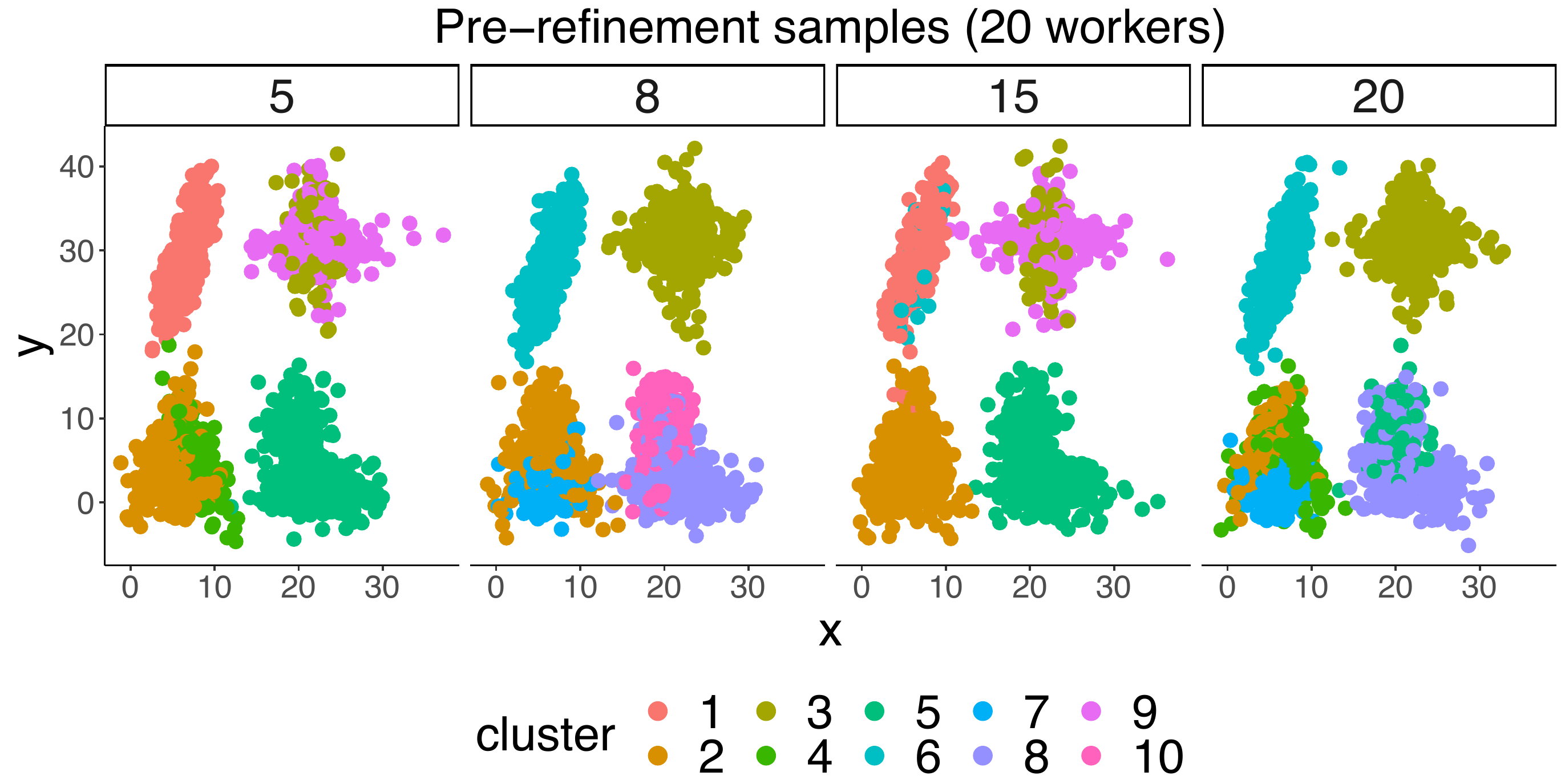}
\caption{Scatter plots of clustering results associated with the training set of one of the synthetic data sets with $N = 10^6$. The plot on the left represents the final clustering estimates when the data are randomly distributed to 1, 5, 20 and 120 workers respectively. The plot on the right shows subsets results associated with the 20 workers on the left; specifically, it represents a sample of cluster allocations of subset 5, 8, 15 and 20 before global cluster refinement. \label{fig:syndat_clustres}}
\end{figure}

Global cluster refinement has benefits beyond label alignment; its general ability to improve the clustering performance for each subset is illustrated by Figure \ref{fig:syndat_accuracydelta} and \ref{fig:syndat_clustres}. Figure \ref{fig:syndat_accuracydelta} illustrates the increase in accuracy associated with the subsets after this procedure. Specifically, for each node setting, we randomly select a run from the ten replicates and draw a violin plot of the changes in accuracy associated with each subset. The before-refinement accuracy is identified by the optimal iteration that minimizes the expected loss locally, whereas the after-refinement accuracy is identified by the final global clustering estimate mapped to local subsets. We notice a general gain in accuracy, which is especially prominent for the flow cytometry data with one million observations.

The benefits from global cluster refinement are also corroborated by Figure \ref{fig:syndat_clustres}, which shows clustering results of selected subsets before refinement (in the bottom row) and the entire data set after refinement (in the top row) using 20 workers. The spurious clusters, coming in various shapes and locations, are identified in local subsets but are eliminated from the final clustering estimate by the global cluster refinement step.

\subsubsection{Classification performance}
\begin{figure}
\centering
	\includegraphics[width =0.7\textwidth]{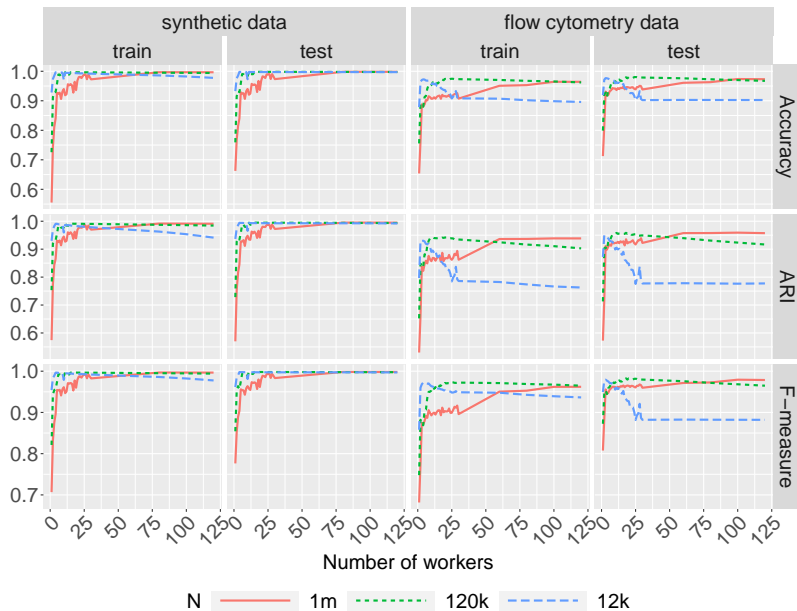}
	\caption{Classification performance of the training and test sets associated with the synthetic (left) and flow cytometry (right) data, as measured by accuracy, ARI, and F-measure.\label{fig:class}}
\end{figure}

\begin{figure}
\includegraphics[width = 0.48\textwidth]{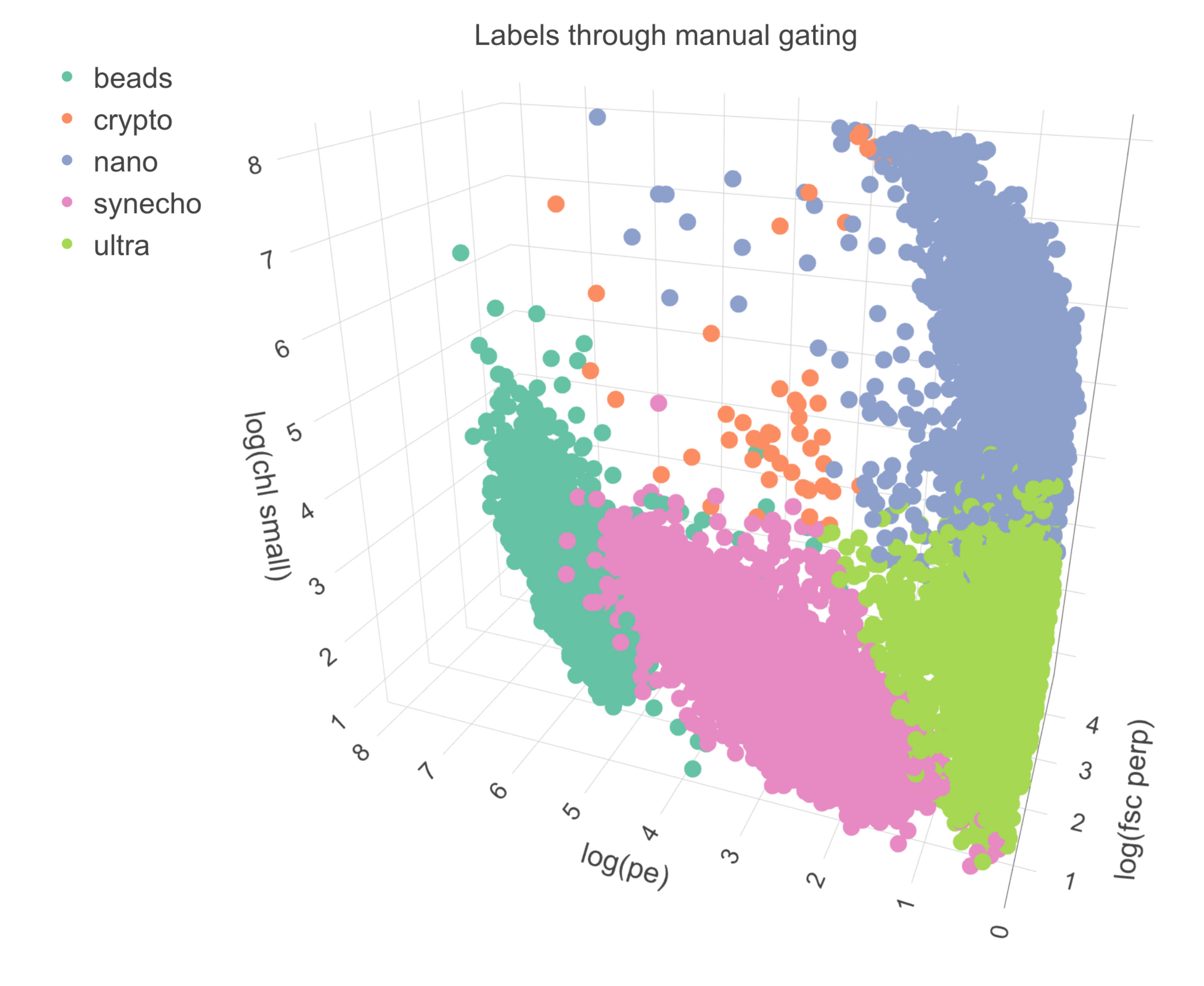}
\includegraphics[width = 0.48\textwidth]{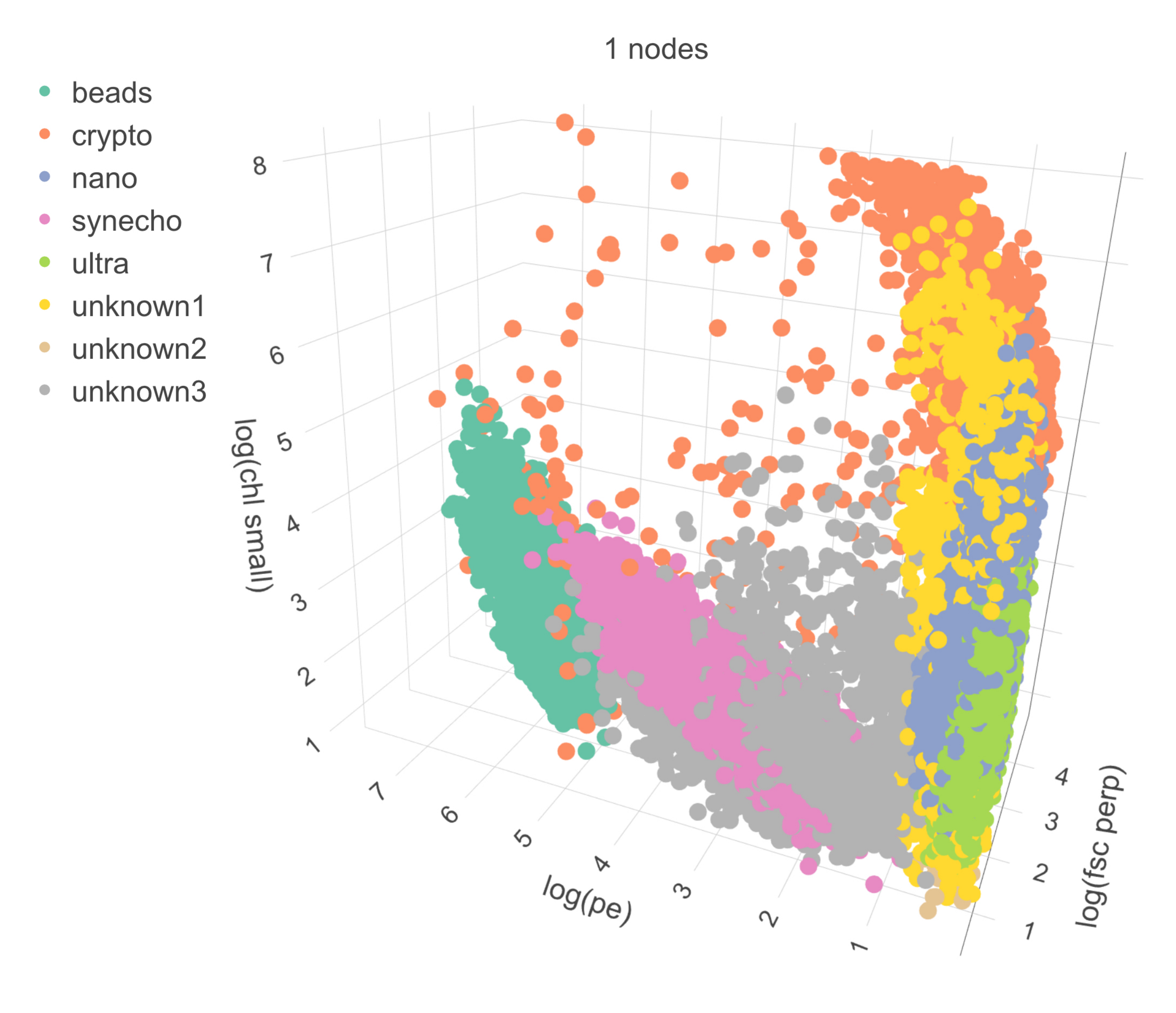}
\includegraphics[width = 0.48\textwidth]{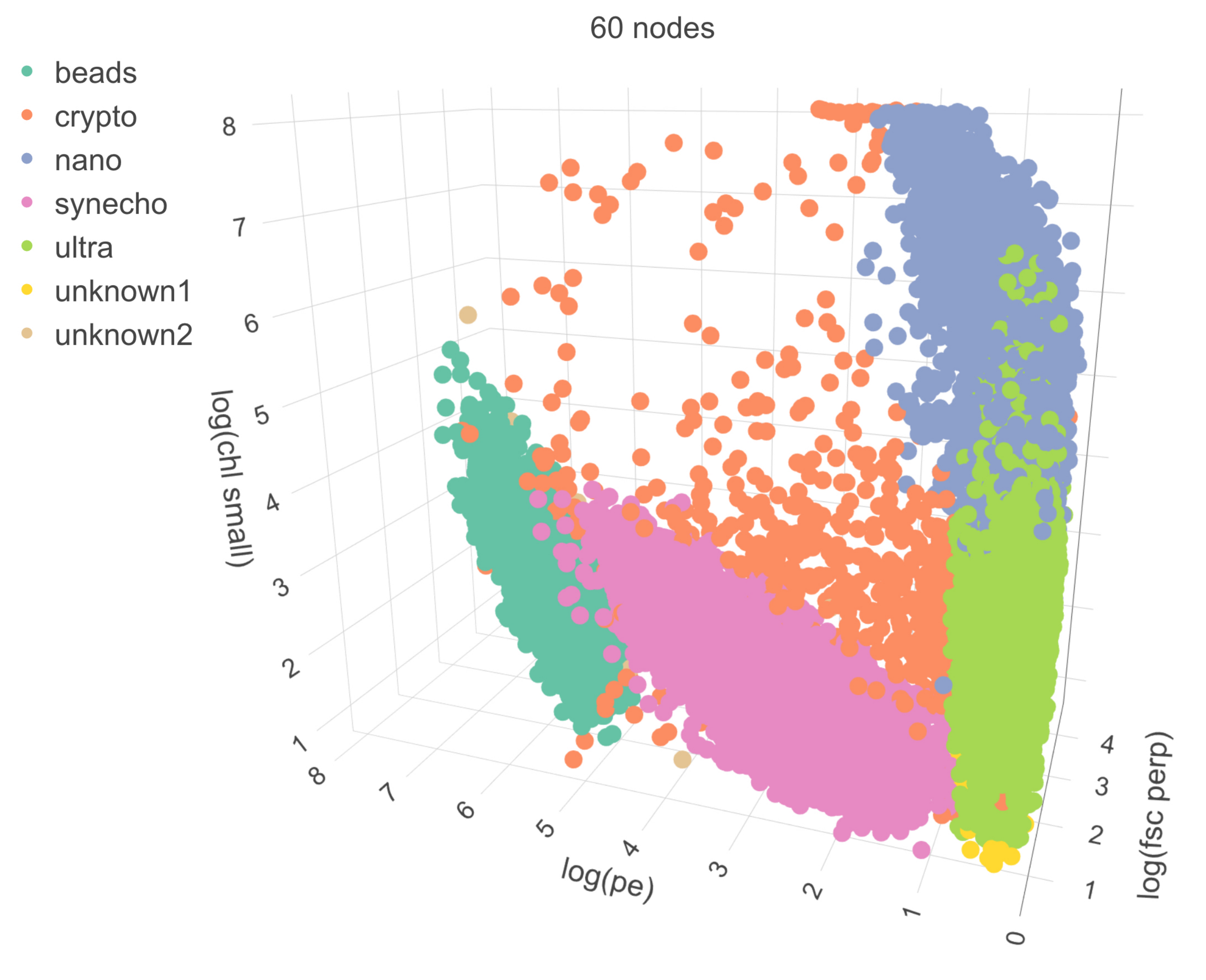}
\includegraphics[width = 0.48\textwidth]{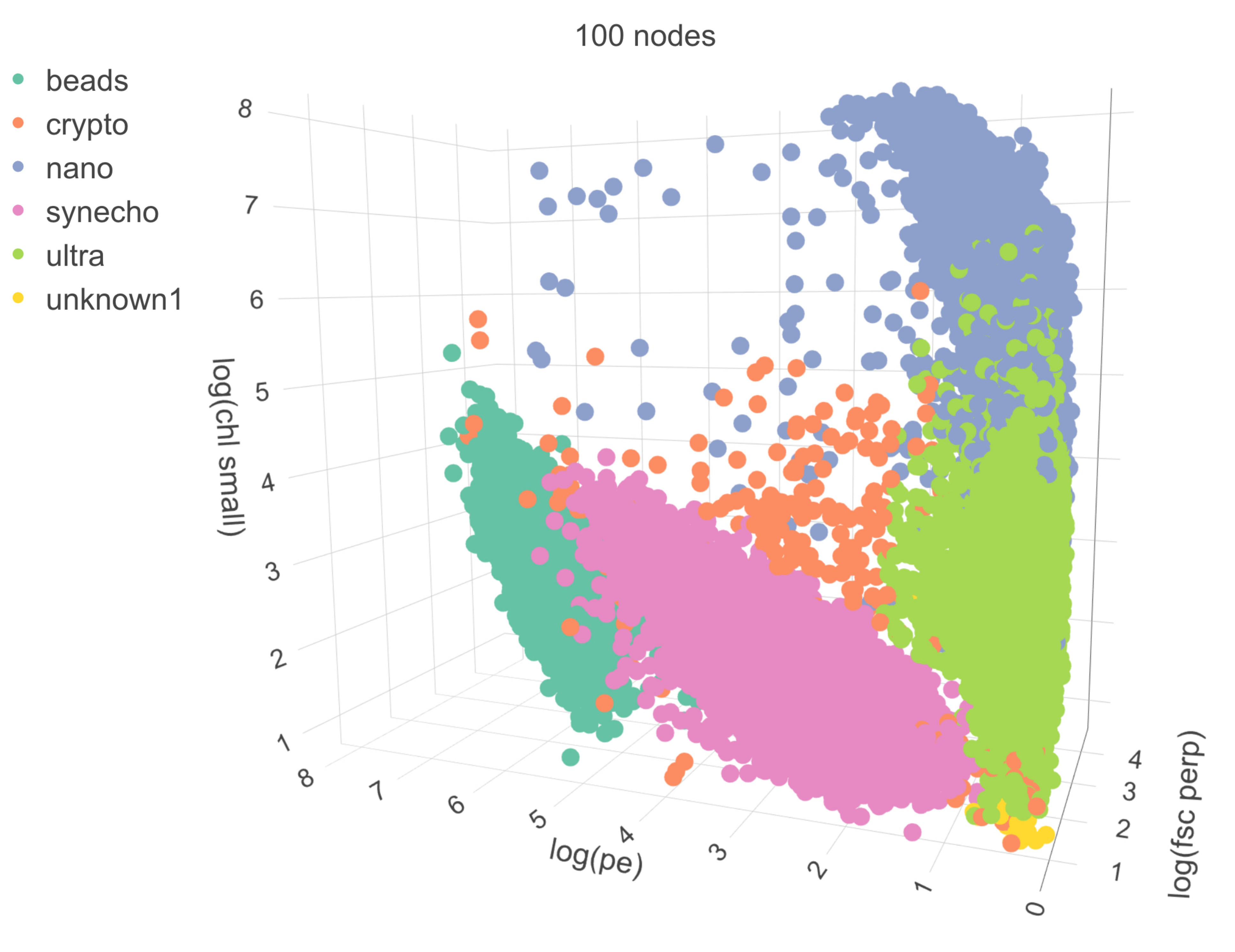}	\caption{Scatter plots of clustering results of the test set associated with $N = 1$ million data size setting. The top left plot corresponds to the manually-gated labels. The remaining four represent the clustering results from using 1, 60 and 100 workers respectively. In each scatter plot, only a random subset of the data points are included; and the $x$, $y$ and $z$ coordinates are $\log$(side scatter), $\log$(phycoerythrin level) and $\log$(chlorophyll level), respectively. These three explain the most variation in the data among the four variables (according to principle component analysis results). \label{fig:seaflow_clustres}}
\end{figure}

DIB-C exhibits promising classification performance, as measured by accuracy, ARI and F-measure, for both training and test sets, with results illustrated in Figure  \ref{fig:syndat_clustres}, \ref{fig:class}, \ref{fig:seaflow_clustres} and Table \ref{tab:comparison}. Figure \ref{fig:class} shows that these measures are robust to the growing number of workers for large (i.e. $N = $ 120k and 1m) data sets; in fact, they increase steadily for the data sets with 1 million observations as the number of workers increases to 120, likely due to the improved mixing of MCMC as the sample size per node drops. This may also explain the sharp rises in performance as we increase the number of workers from one to two. These observations indicate a win-win position: DIB-C is not only scalable, but also robust to the increasing number of workers. 

Figure \ref{fig:syndat_clustres} and \ref{fig:seaflow_clustres} show the scatter plots of the clustering results associated with a training set of the synthetic data and a test set of the flow cytometry data, respectively, when $N = 1$ million. As evidenced by both figures, the clustering tends to be less noisy and visually more reasonable as the number of workers increases. In Figure \ref{fig:seaflow_clustres}, the names of the species generally match their manually-gated labels when the number of workers is large, with synecho---which is disconnected in the manually-gated labels---decomposed into two clusters.

We have also compared DIB-C with  popular distributed clustering algorithms, including DBSCAN (density-based spatial clustering of applications with noise) and K-means. In terms of the tuning parameters, for DBSCAN, we set $\epsilon$ based on a widely used k-nearest neighbour distance plot and \emph{minPts} at the value that maximizes the accuracy, ARI and F-measure for the training set (they all happen to be maximized by the same minPts); for K-means, we set the number of clusters to be 5 (the truth according to the manual gating), 6 and 8 respectively. The distributed version of DBSCAN and K-means are much faster than DIB-C and yield the same clustering performance as their non-distributed counterparts.
However, their performance, as shown in Table \ref{tab:comparison}, is very sensitive to tuning parameters; taking K-means as an example, the slight deviation of $K$ from the truth leads to drastically worse performance. In addition, DIB-C, for which the median performance statistics are reported, consistently outperforms all methods with the exception of K-means when the number of clusters is correctly specified. However, even K-means with the number of clusters correctly specified cannot provide a generative model, based on which we can simulate similar data and perform density estimation and other inference. Our method, however, enjoys such an advantage, as is shown in Section \ref{sec:postpred}.

We also attempted comparison with Dirichlet-process Gaussian mixtures, another widely used model-based approach for clustering. Since the computation was already extremely slow for $N = 12k$ (taking over 20 hours), we dropped the comparison.

\begin{table}[]
\caption{\label{tab:comparison} The clustering performance of the flow cytometry data across various methods for $N = $ 12k, 120k and 1m. In particular, the performance of DBSCAN and K-means are invariant to the number of workers. For DIB-C, the median of each statistic obtained from using varying number of workers is reported. Yellow and green indicate the best and the second best performance under each scenario respectively. }
\resizebox{\columnwidth}{!}{%
\begin{tabular}{cl|ccc|ccc}
\multicolumn{1}{l}{}  &                              & \multicolumn{3}{c|}{train}   & \multicolumn{3}{c}{test}    \\
\multicolumn{1}{l}{} &                              & Accuracy & ARI   & F-measure & Accuracy & ARI   & F-measure \\ \hline
\multirow{7}{*}{12k}  & DIB-C                        & \cellcolor{green} 0.938 & \cellcolor{green} 0.849 & \cellcolor{green}0.955    &\cellcolor{green} 0.939    &\cellcolor{green} 0.860 & \cellcolor{green}0.933     \\
                      & DBSCAN- default              & 0.719    & 0.350 & 0.691     & 0.716    & 0.352 & 0.687     \\
                      & \bigcell{c}{DBSCAN-optimal \tabularnewline (minPts = 40)} & 0.909    & 0.803 & 0.895     & 0.906    & 0.790 & 0.882     \\
                      & K-means, $K = 5$             & \cellcolor{yellow}0.977    & \cellcolor{yellow}0.945 & \cellcolor{yellow}0.980     &  \cellcolor{yellow}0.976    &  \cellcolor{yellow}0.941 &  \cellcolor{yellow}0.979     \\
                      & K-means, $K = 6$             & 0.699    & 0.593 & 0.796     & 0.714    & 0.603 & 0.809     \\
                      & K-means, $K = 8$             & 0.621    & 0.537 & 0.748     & 0.638    & 0.547 & 0.761     \\ \hline
\multirow{7}{*}{120k} & DIB-C                        &\cellcolor{green}0.966
 & \cellcolor{green}0.938 & \cellcolor{green}0.971    & \cellcolor{green}0.965    & \cellcolor{yellow}0.952 & \cellcolor{green}0.976     \\
                      & DBSCAN-default               & 0.904    & 0.793 & 0.886     & 0.903    & 0.792 & 0.885     \\
                      &  \bigcell{c}{DBSCAN-optimal \tabularnewline (minPts = 5)}  & 0.904    & 0.793 & 0.886     & 0.904    & 0.792 & 0.884     \\
                      & K-means, $K = 5$             & \cellcolor{yellow}0.979    &\cellcolor{yellow} 0.950 & \cellcolor{yellow}0.982     &\cellcolor{yellow} 0.979    & \cellcolor{green}0.950 & \cellcolor{yellow}0.982     \\
                      & K-means, $K = 6$             & 0.706    & 0.706 & 0.803     & 0.707    & 0.707 & 0.802     \\
                      & K-means, $K = 8$             & 0.625    & 0.543 & 0.751     & 0.625    & 0.546 & 0.70     \\ \hline
\multirow{7}{*}{1m}  & DIB-C                       &   \cellcolor{green} 0.913    &  \cellcolor{green}0.868 &  \cellcolor{green}0.901     &  \cellcolor{green}0.943    &  \cellcolor{green}0.924& \cellcolor{green} 0.962     \\
                      & DBSCAN-default               & 0.718    & 0.356 & 0.692     & 0.718    & 0.356 & 0.691     \\
                      &  \bigcell{c}{DBSCAN-optimal \tabularnewline (minPts = 12)} & 0.904    & 0.791 & 0.885     & 0.904    & 0.791 & 0.885     \\
                      & K-means, $K = 5$             & \cellcolor{yellow} 0.979    &\cellcolor{yellow} 0.949 &\cellcolor{yellow} 0.982     & \cellcolor{yellow}0.978    & \cellcolor{yellow}0.948 &\cellcolor{yellow} 0.981     \\
                      & K-means, $K = 6$             & 0.708    & 0.599 & 0.803     & 0.708    & 0.599 & 0.804     \\
                      & K-means, $K = 8$             & 0.630    & 0.544 & 0.755     & 0.630    & 0.544 & 0.755     \\ \hline
\end{tabular}%
}
\end{table}

\subsubsection{Simulation from posterior predictive\label{sec:postpred}}

\begin{figure}
    \includegraphics[width = 0.5\textwidth]{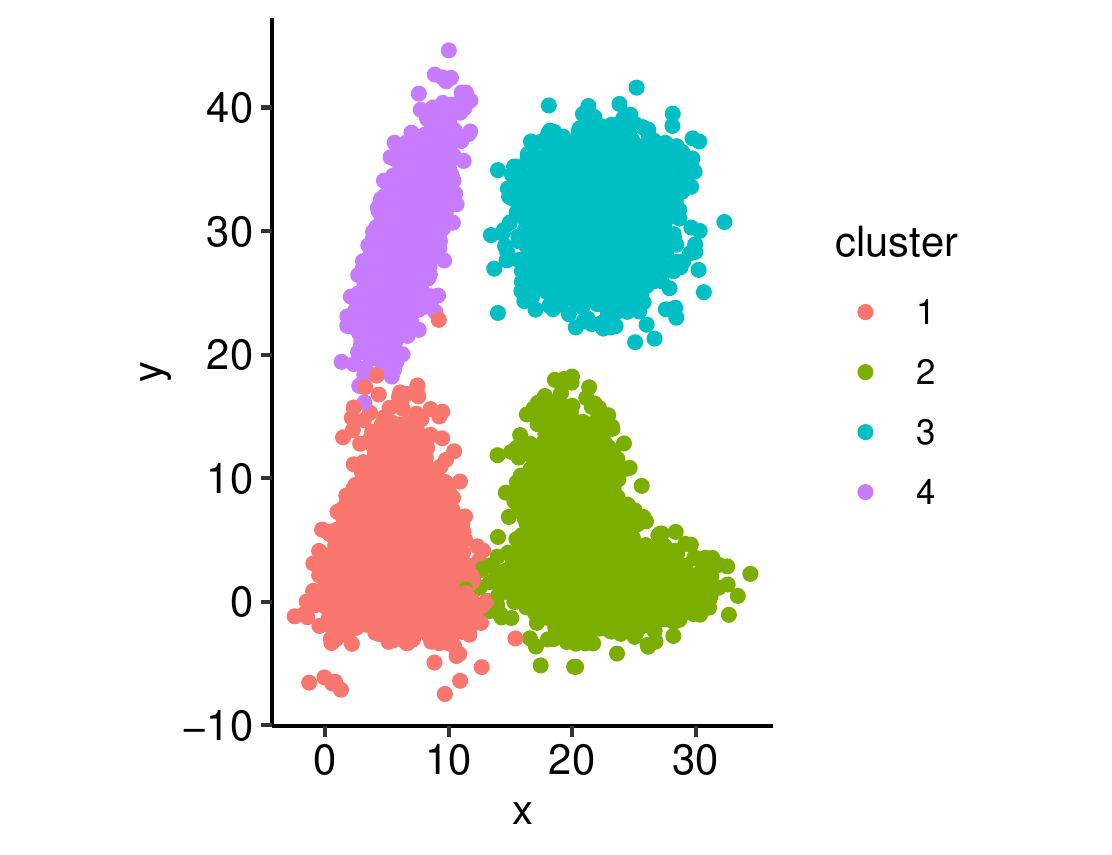}
    \includegraphics[width =0.5\textwidth]{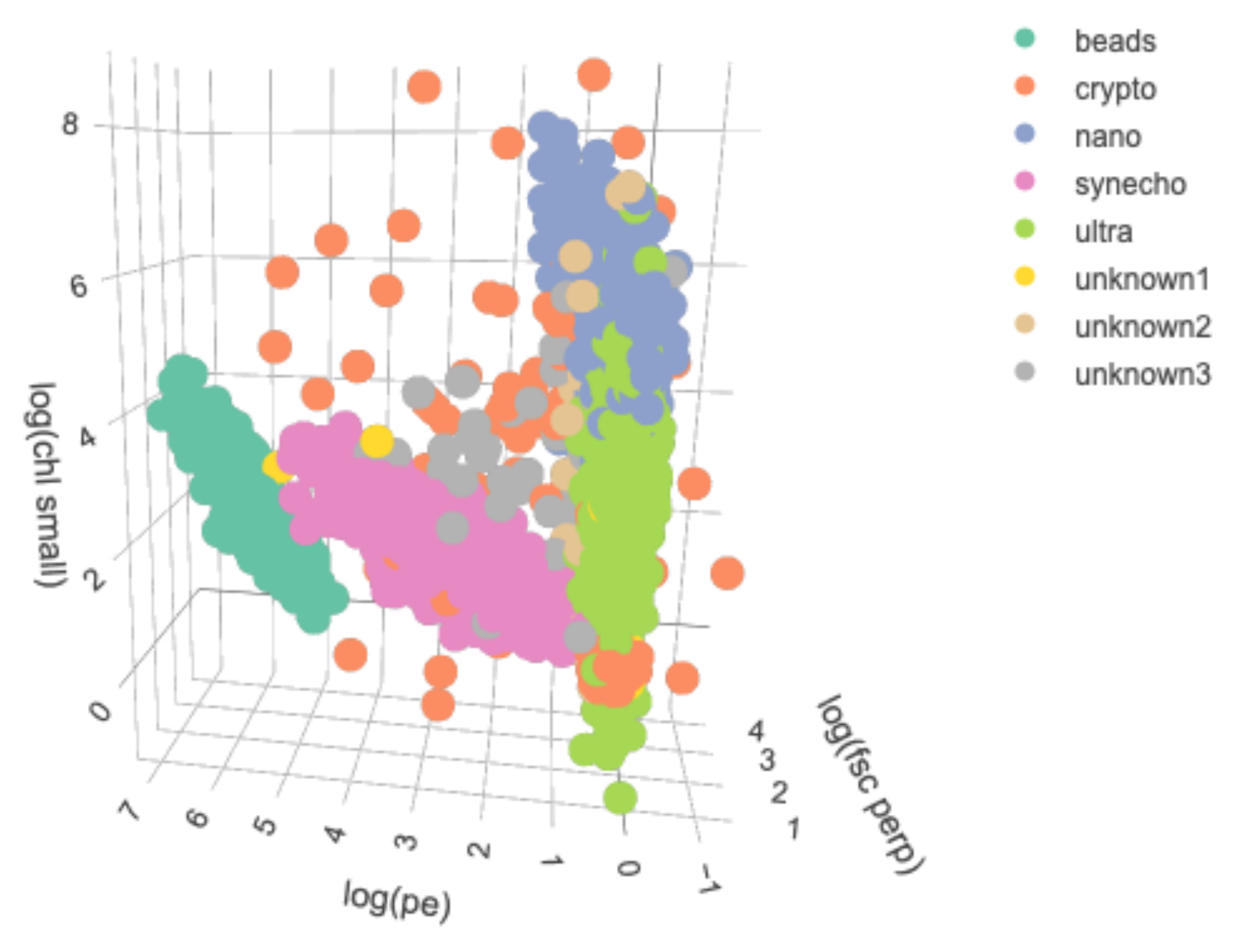}
    \caption{Scatter plots of 10,000 data points simulated from the posterior predictive distributions resulting from an analysis conducted with data distributed across 40 workers for the synthetic data (left) and the flow cytometry data (right). In the plot on the right, chl small, pe and fsc perp represent chlorophyll level, phycoerythrin level, and forward scatter respectively.}
    \label{fig:post_pred}
\end{figure}

Figure \ref{fig:post_pred} shows scatter plots of 10,000 data points simulated from the posterior predictive distributions resulting from analyses in which the synthetic and the flow cytometry data are randomly distributed across 40 workers. The posterior predictive distributions show striking resemblance to the true data; the cluster labels, created by finding a map that minimizes the number of mismatches to the true or manually-gated labels, also largely match those in the original data set.

\begin{figure}
	\centering
		\includegraphics[width =0.7 \textwidth]{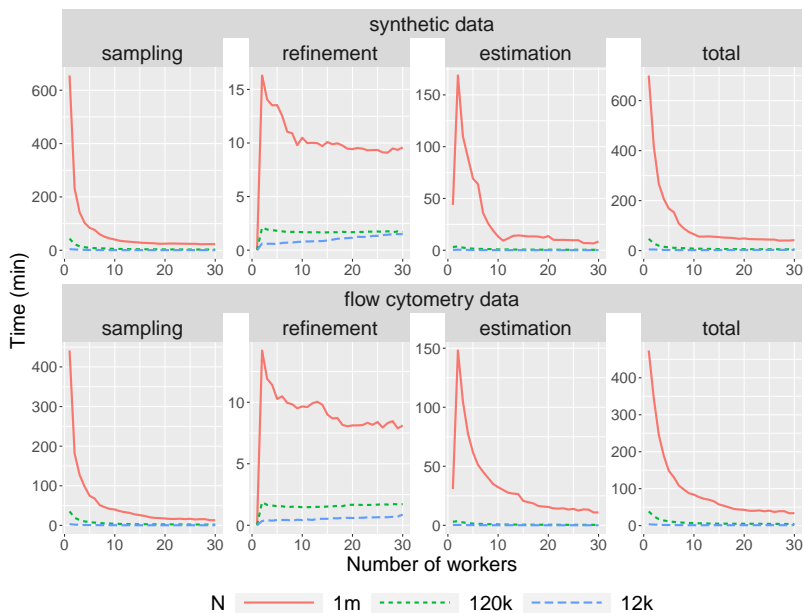}
	\caption{The computation time associated with the training sets of the synthetic (top) and flow cytometry data (bottom). For each data setting, we include sampling local clustering (the first column), global cluster refinement (the second column), global clustering estimation time (the third column) and total time of the above steps (the fourth column). \label{fig:syndat_time}}
\end{figure}
\subsubsection{Computation time}
Figure \ref{fig:syndat_time} displays the computation time for sampling local clustering, global cluster refinement, global clustering estimation and the total run time, respectively. We notice similar trends in both data settings, most notably the drastic decrease in the total computation time of the data sets with one million observations; the dive, quickly plateauing at around 15 workers, is mostly attributable to a significant speedup, dropping from around 8.5 hours to 15 minutes, from parallel MCMC in sampling local clustering. No global cluster refinement or global clustering estimation is necessary in a conventional single machine scenario, which explains the jump in the associated time when switching to two workers. 

\section{Conclusion}
In this paper, we have proposed a nearly embarrassingly parallel framework named DIB-C for distributed Bayesian clustering under a finite mixture of Gaussian mixtures model. DIB-C accommodates any loss function on the space of partitions for cluster estimation, quickly classifies future subjects, and allows density estimation and other posterior inference. Our extensive experiments also demonstrate that DIB-C is not only scalable, but also robust to the increasing number of workers for large data sets.

\acks{The authors acknowledge support from SAS Institute Inc., United States National Science Foundation grant 1546130 and National Institute of Environmental Health Sciences of the National Institutes of Health grant R01ES027498.}

\clearpage
%\newpage
\appendix

% Note: in this sample, the section number is hard-coded in. Following
% proper LaTeX conventions, it should properly be coded as a reference:

%In this appendix we prove the following theorem from
%Section~\ref{sec:textree-generalization}:

%\section*{Appendix A.}
\section{MCMC procedures for sampling of local clustering (step 2)\label{appendix:mcmc}}

We adapt \citet{Wallietal17}'s MCMC sampling scheme for finite mixture of mixtures for sampling of local clustering (step 2) and model parameters (step 5) in DIB-C. The sampling scheme is based on a block conditional Gibbs sampler with data augmentation. In this section, we introduce procedures for sampling of local clustering (step 2).

Assume $\mathcal{Y}$ can be partitioned into $R$ non-overlapping subsets, with subset $r$, denoted by $\mathcal{Y}_r$, residing on worker $r$ $(r = 1, \ldots, R)$. We run MCMC on $R$ workers in an embarrassingly parallel manner, producing draws from subset posterior $p_r(\Theta, \eta \mid \mathcal{Y}_r)$, $r = 1, \ldots, R$. Without loss of generality, we introduce the sampling for subset $r$. Recall the definition of subset posterior $r$
\begin{align*}
p_r(\Theta, \eta \mid\mathcal{Y}_r) &= \frac{\{\prod_{i = 1}^{n_r} f(y_{ri} \mid \Theta, \eta)\}p(\Theta, \eta)}{\int \int \{\prod_{i = 1}^{n_r} f(y_{ri} \mid \Theta, \eta)\}p(\Theta, \eta)d\Theta d \eta}.
\end{align*} 

%in which subset likelihood is raised to the power of $a := N / N_r \approx R$ to ensure that the variance of each subset posterior is roughly on the same scale as that of the overall posterior. 
Recall $\mathbf{c}_r = (c_{r1}, \ldots, c_{rn_r})$ is the vector of latent cluster allocations, which take values  in $\{1, \ldots, K\}^{n_r}$, indicating the cluster to which each observation belongs such that\begin{align*}p(y_{ri} \mid \theta_1, \ldots, \theta_K, c_{ri}= k ) = p_k(y_{ri} \mid \theta_k), \quad \text{and}\quad P\{c_{ri} = k \mid \eta\} = \eta_k.\end{align*} Recall the vector of latent subcomponent allocations $\mathbf{s}_r = (s_{r1}, \ldots, s_{ rn_r})$, which take values in $\{1, \ldots, L\}^{n_r}$, to indicate the subcomponent to which an observation within a cluster is assigned such that 
	\begin{align*}p_k(y_{ri}\mid\theta_k, c_{ri} = k, s_{ri} = l) = f_{\text{N}}(y_{ri} \mid \mu_{kl}, \Sigma_{kl}), \quad\text{and}\quad P\{s_{ri} = l \mid c_{ri} = k, w_k\} = \omega_{kl}.
	\end{align*}

%Let $\varphi_0 = (e_0, d_0, c_0, g_0, \mathbf{G}_0, \mathbf{B}_0, \mathbf{m}_0, \mathbf{M}_0,\nu)$ be a set of fixed hyper-parameters. 
Using the priors specified in Section \ref{sec:model}, the sampling steps are given as follows:
%\mathbf{c}_r, \mathbf{s}_r)$ and parameters $(\eta, \omega_k, \mu_{kl}, \Sigma_{kl}, C_{0k}, b_{0k}, \lambda_{kj})$, $k = 1, \ldots, K$, $l = 1, \ldots, L$, $j = 1, \ldots, d$, are sampled from the {subset posterior} using the following Gibbs sampling scheme. The sampler alternates between updating cluster weights $\eta$, imputing latent cluster allocation $\mathbf{c}_r$, and updating cluster-specific parameters, which constitutes sampling for subcomponents within each cluster.
		\begin{enumerate}
		\item \textbf{Sampling steps on the level of the cluster distribution:}
		\begin{enumerate}[label=\textbf{A.\arabic*}]
			\item{\label{item:eta}} Parameter simulation step conditional on the classification \\  $\mathbf{c}_r$. Sample $\eta\mid\mathbf{c}_r$ from $\text{Dir}(e_1, \ldots, e_K)$, $e_k = e_0 + \mathfrak{n}_k$, $k = 1, \ldots, K$, where $\mathfrak{n}_k =  \#\{i: c_{ri} = k\}$ is the number of observations allocated to cluster $k$.
			\item Classification step for each observation $y_{ri}$ conditional on cluster-specific parameters. \\
			For each $i = 1, \ldots, n_r$ sample the cluster assignment $c_{ri}$ from
			\begin{align*}
			P\{c_{ri} = k \mid y_{ri},\theta,\eta\} \propto \eta_k p_k(y_{ri} \mid \theta_k), k = 1, \ldots, K,
			\end{align*}
			where $p_k(y_{ri} \mid\theta_k)$ is the semi-parametric mixture approximation of the cluster density: 
			\begin{align*}
			p_k(y_{ri} \mid \theta_k) = \sum_{l = 1}^L w_{kl}f_N(y_{ri} \mid \mu_{kl}, \Sigma_{kl}).
			\end{align*}
		%	Note that clustering of the observations is performed on the upper level of the model, using a collapsed Gibbs step, where the latent, within-cluster allocation variables $\mathbf{I}$ are intergrated out.
		\end{enumerate}
		\item \textbf{Within each cluster $k$, $k = 1, \ldots, K$:}
		\begin{enumerate}[label=\textbf{B.\arabic*}]
			\item Classification step for all observations $y_{ri}$ that are assigned to cluster $k$ (i.e. $c_{ri} = k$), conditional on the subcomponent weights and the subcomponent-specific parameters.\\ For each $i \in \{i = 1, \ldots, n_r: c_{ri} = k\}$ sample $s_{ri}$ from 
			\begin{align*}
			Pr\{s_{ri} = l \mid y_{ri}, \theta_k, c_{ri} = k\} \propto w_{kl}f_\text{N}(y_{ri} \mid \mu_{kl}, \Sigma_{kl}), \,l = 1,\ldots, L.
			\end{align*}
			\item{\label{item:subcomp}} Parameter simulation step conditional on the classifications $\mathbf{s}_r$ and $\mathbf{c}_r$:
			\begin{enumerate}
				\item Sample $w_k \mid \mathbf{c}_r, \mathbf{s}_r$ from $\text{Dir}(d_{k1}, \ldots, d_{kL})$, $d_{kl} = d_0 + 
				\mathfrak{n}_{kl}$, $l = 1,\ldots, L$, where $\mathfrak{n}_{kl} =  \#\{i: s_{ri} = l, c_{ri} = k\}$ is the number of observations allocated to subcomponent $l$ in cluster $k$.	
				\item For $l= 1,\ldots,L$: Sample $\Sigma_{kl}^{-1}\mid\mathbf{c}_r,\mathbf{s}_r, \mu_{kl}, C_{0k},\mathcal{Y}_r \sim \text{W}_d(c_{kl}, C_{kl})$,where
				\begin{align*}	c_{kl} &= c_0 +  \mathfrak{n}_{kl} , \\
				C_{kl} &= C_{0k} +   \sum_{i:s_{ri} = l,c_{ri} = k} (y_{ri} - \mu_{kl})(y_{ri} - \mu_{kl})^T
				\end{align*}
				\item For $l= 1,\ldots, L$: Sample $\mu_{kl}\mid\mathbf{c}_r,\mathbf{s}_r,b_{0k}, \Sigma_{kl}, \Lambda_{k}, \mathcal{Y}_r \sim \text{N}_d\left(b_{kl}, B_{kl}\right)$, where \begin{align*}
				B_{kl} &= \left(\tilde{B}_{0k}^{-1} + \mathfrak{n}_{kl} \Sigma_{kl}^{-1}\right)^{-1},\\
				b_{kl} &= B_{kl}\left(\tilde{B}_{0k}^{-1}b_{0k} + \Sigma_{kl}^{-1}  n_{kl} \bar{y}_{kl}\right),
				\end{align*}
				with $\tilde{B}_{0k} = \sqrt{\Lambda_k}B_0 \sqrt{\Lambda_k}$, $\Lambda_k = diag(\lambda_{k1}, \ldots, \lambda_{kd})$, and\\ $\bar{y}_{kl}\mathfrak{n}_{kl} =  \sum_{i,s_{ri} = l, c_{ri} = k}y_{ri}.$
			\end{enumerate}
		\end{enumerate}
		\item{\label{item:hyperpara}} \textbf{For each cluster $k, k = 1,\ldots,K$: Sample the random hyperparameters $\lambda_{kj}, C_{0k},b_{0k}$ from their full conditionals:}
		\begin{enumerate}[label=\textbf{C.\arabic*}]
			\item For $j = 1, \ldots, d$: Sample $\lambda_{kj} \mid b_{0k}, \mu_{k1},\ldots, \mu_{kL} \sim \text{GIG}(p_{kL}, a_{kj}, b_{kj})$, where $\text{GIG}$ is the generalized inverse Gaussian distribution and \begin{align*}
			p_{kL} &= -L/2 + \nu,\\a_{kj} &= 2\nu,\\b_{kj} &=\sum_{l = 1}^L(\mu_{kl,j} - b_{0k,j})^2/B_{0, jj}.
			\end{align*}
			\item Sample $C_{0k}\mid \Sigma_{k1}, \ldots, \Sigma_{kL} \sim \text{W}_d\left(g_0 + Lc_0, G_0 + \sum_{l =1}^L\Sigma_{kl}^{-1}\right)$.
			\item Sample $b_{0k}\mid\tilde{B}_{0k},\mu_{k1}, \ldots, \mu_{kL} \sim \text{N}_d\left(\tilde{m}_k, \tilde{M}_k\right)$, where
			\begin{align*}
			\tilde{M}_k &= \left(M_0^{-1} + L\tilde{B}_{0k}^{-1}\right)^{-1},\\
			\tilde{m}_k &= \tilde{M}_k \left(M_0^{-1}m_0 + \tilde{B}_{0k}^{-1}\sum_{l = 1}^L \mu_{kl}\right).
			\end{align*}
		\end{enumerate}
	\end{enumerate}
%	This algorithm applies to both semi-supervised and unsupervised clustering problem. A semi-supervised learning problem contains partially missing cluster assignments, so only unlabeled data need to update their cluster allocations at every MCMC iteration. Since labeled data points fixate the cluster allocation, there is no label switching problem locally and clusters labels across workers can be automatically aligned.
%For sampling of local clustering (step 2), one is only required to produce posterior imputation of local cluster and subcomponent allocations. Therefore, one can use any other sampling scheme to improve mixing (such as a collapsed Gibbs sampler) as long as samples of local cluster and subcomponent allocations are produced. 

\section{Collapsed Gibbs sampler for global cluster refinement\label{appendix:collapsed}}
The key quantity in updating the group allocation is the posterior probability of item $b$ being assigned to group $h$: 
\begin{align}P(\mathscr{z}_b = h \mid \mathbf{\mathbf{z}}_{\setminus{b}}, \mathcal{Y}) &\propto P(\mathscr{z}_{b} = h \mid \mathbf{z}_{\setminus{b}})p(\mathcal{Y} \mid \mathscr{z}_b = h, \mathbf{z}_{\setminus b}) \nonumber \\ 
&= P(\mathscr{z}_b = h \mid \mathbf{z}_{\setminus{b}}) p(\mathscr{y}_b \mid \mathcal{Y}_{\setminus b}, \mathscr{z}_b = h, \mathbf{z}_{\setminus{b}}) p(\mathcal{Y}_{\setminus b}\mid \mathscr{z}_b = h, \mathbf{z}_{\setminus{b}})\nonumber \\
&\propto P(\mathscr{z}_b = h \mid \mathbf{z}_{\setminus{b}}) p(\mathscr{y}_b \mid \mathcal{Y}_{\setminus b}, \mathscr{z}_b = h, \mathbf{z}_{\setminus{b}}), \label{equ:post_collapsedGibbs}\end{align}
where $h = 1, \ldots, H; b = 1, \ldots, B.$
Note that \begin{align*}P(\mathbf{z}) = \frac{\Gamma(H\alpha_0)}{\Gamma(N + H\alpha_0)} \prod_{h = 1}^H \frac{\Gamma{(\mathscr{N}_h + \alpha_0)}}{\Gamma(\alpha_0)},\end{align*}
so \begin{align}P(\mathscr{z}_b = h, \mathbf{z}_{\setminus{b}}) &=
\frac{\Gamma(H\alpha_0)}{\Gamma(N + H\alpha_0)} \frac{\Gamma{(\mathscr{N}_{h\setminus{b}} + \mathscr{n}_b + \alpha_0)}}{\Gamma(\alpha_0)}\prod_{j = 1, j \neq h}^H \frac{\Gamma{(\mathscr{N}_j + \alpha_0)}}{\Gamma(\alpha_0)}, \label{equ:term11}\\
P(\mathbf{z}_{\setminus{b}})&= \frac{\Gamma(H\alpha_0)}{\Gamma(N  - \mathscr{n}_b + H\alpha_0)} \frac{\Gamma{(\mathscr{N}_{h \setminus{b}} + \alpha_0)}}{\Gamma(\alpha_0)}\prod_{j = 1, j \neq h}^H \frac{\Gamma{(\mathscr{N}_j + \alpha_0)}}{\Gamma(\alpha_0)}\label{equ:term12}
\end{align} hold.
By \eqref{equ:term11} and \eqref{equ:term12}, the first term in \eqref{equ:post_collapsedGibbs} is
\begin{align*}
P(\mathscr{z}_{b} = h \mid \mathbf{z}_{\setminus{b}}) &= \frac{P(\mathscr{z}_b = h, \mathbf{z}_{\setminus{b}})}{P(\mathbf{z}_{\setminus{b}})}\\
&=\frac{\Gamma(N + H\alpha_0 - \mathscr{n}_b )\Gamma(\mathscr{N}_{h \setminus{b}} + \mathscr{n}_b + \alpha_0) }{\Gamma(N + H\alpha_0) \Gamma(\mathscr{N}_{h \setminus{b}} + \alpha_0)}.
\end{align*}
Let ${y}^*$ be a new data vector. \begin{align}
    p({y}^* \mid \mathcal{Y}) = \text{t}_d\left({y}^*\mid \mathbf{m}_N, \frac{\kappa_N + 1}{\kappa_N(\nu_N - d + 1)}\mathbf{S}_N, \nu_N - D + 1\right),\label{equ:term2}
\end{align}
where $\mathbf{m}_N = {N \bar{{y}}}/{\kappa_N}$, $\kappa_N = 1 + N,$ $\nu_N = \nu_0 + N,$ and $\mathbf{S}_N = \mathbf{S}_0 + \sum_{n=1}^N {y}_n {y}_n^T - \kappa_N \mathbf{m}_N\mathbf{m}_N^T.$
The second term in \eqref{equ:post_collapsedGibbs} can be re-expressed as
\begin{align*}
    p(\mathscr{y}_b \mid \mathcal{Y}_{\setminus{b}}, \mathscr{z}_b = h, \mathbf{z}_{\setminus{b}}) = p(\mathscr{y}_b \mid \mathcal{Y}_{h \setminus{b}}),
\end{align*}
which is given by a product of $t$-densities based on \eqref{equ:term2}
\begin{align*}
p(\mathscr{y}_b \mid \mathscr{Y}_{h \setminus{b}}) = \prod_{j = 1}^{\mathscr{n}_b} \text{t}_d\left(\mathscr{y}_{b,j}\,\Bigg|\, \mathbf{m}_{h \setminus{b}},  \frac{\kappa_{h \setminus{b}} + 1}{\kappa_{h \setminus{b}}(\nu_{h \setminus{b}}- d + 1) }\mathbf{S}_{h \setminus{b}}, \nu_{h \setminus{b}} - d + 1\right),
\end{align*}
where \begin{align*}\kappa_{h \setminus{b}} &= 1 + \mathscr{N}_{h \setminus{b}}, \quad \nu_{h \setminus{b}} = \nu_0 + \mathscr{N}_{h \setminus{b}}, \\ \mathbf{m}_{h \setminus{b}} &= \frac{\mathscr{N}_{h \setminus{b}} \bar{\mathscr{Y}}_{h \setminus{b}}}{\kappa_{h \setminus{b}}} \quad\text{and}\quad \mathbf{S}_{h \setminus{b}} = \mathbf{S}_0 + \mathscr{N}_{h \setminus{b}}\mathscr{S}_{h \setminus{b}} - \kappa_{h \setminus{b}} \mathbf{m}_{h \setminus{b}} \mathbf{m}_{h \setminus{b}}^T. \end{align*}

\section{Sampling model parameters (step 5) \label{appendix:alg_sampling}}
Algorithm \ref{alg:sampling} outlines the steps in approximate posterior sampling of model parameters (step 5).
\begin{algorithm}
   	\caption{Sampling Model Parameters}
	% \newline(Parallel Collapsed Gibbs sampler for a Gaussian mixture model)
	\label{alg:sampling}
	\hspace*{\algorithmicindent} \textbf{Input:} Output of refined samples of local cluster allocations from Algorithm \ref{alg:align}\\
 \hspace*{\algorithmicindent} \textbf{Output:} Optimal clustering estimate $\mathbf{c}^*$
	\begin{algorithmic}[1]
	\State \textbf{On Workers:}
	\ParFor{Worker $r = 1$ to $R$}
	\For{$k = 1$ to $k_N$}
	\For{$l = 1$ to $L$}
	\State $\mathfrak{n}_{rkl} \gets \sum_{i:y_{ri} \in \mathcal{Y}_r} \mathbbm{1}_{\tilde{s}_{ri} = l, \tilde{c}_{ri} = k}$
	\State$\mathfrak{S}_{rkl} \gets \sum_{\substack{y_{ri} \in \mathcal{Y}_r, \\\tilde{s}_{ri} = l, \tilde{c}_{ri} = k}}  y_{ri} y_{ri}^{T}$ 
	\State $\mathscr{s}_{rkl} \gets \sum_{\substack{y_{ri} \in \mathcal{Y}_r, \\\tilde{s}_{ri} = l, \tilde{c}_{ri} = k}} y_{ri}$
\State Send $\mathfrak{n}_{rkl}, \mathfrak{S}_{rkl}$ and $\mathscr{s}_{rkl}$ to Master
\EndFor
\EndFor
\EndParFor
\State \textbf{On Master:}
\For{$k = 1$ to $k_N$}
\For{$l = 1$ to $L$}
\State $\mathfrak{n}_{kl} \gets \sum_{r} \mathfrak{n}_{rkl}$,\quad$\mathfrak{S}_{kl} \gets \sum_{r} \mathfrak{S}_{rkl}$ \quad and \quad $\mathfrak{s}_{kl} \gets \sum_{r} \mathfrak{s}_{rkl}$
\EndFor
\State $\mathfrak{n}_k \gets \sum_{l}\mathfrak{n}_{kl}$
\EndFor
\State Initialize $(\Theta^{(0)}, \eta^{(0)})$  \Comment{Start of the sampling; see Appendix \ref{appendix:mcmc} for more details}
\For{Iteration $t = 1$ to $\mathfrak{T}$}
\State Sample $\eta^{(t)}$ from $\text{Dir}\left(e_1, \ldots, e_{k_N}\right),$ where $e_k = e_0 + \mathfrak{n}_k$\Comment{See Appendix \ref{appendix:mcmc} \ref{item:eta}}
\For{$k = 1$ to $k_N$} 
\LineComment{See Appendix \ref{appendix:mcmc} \ref{item:subcomp}}
\State Sample $\omega_k^{(t)}$ from $\text{Dir}(d_{k1}, \ldots, d_{kL}),$ where $d_{kl} = d_0 + \mathfrak{n}_{kl}$
\For{$l = 1$ to $L$}
\State Sample ${(\Sigma_{kl}^{-1})}^{(t)}$ from $\text{W}_d(c_{kl}, C_{kl}),$ where $c_{kl} = c_0 + \mathfrak{n}_{kl}$, and $C_{kl} = C_{0k}^{(t-1)} + \mathfrak{S}_{kl} - 2\mu_{kl} \mathfrak{s}_{kl} + \mathfrak{n}_{kl}\mu_{kl}^{(t-1)}{(\mu_{kl}^{(t-1)})}^T$
\State Sample $\mu_{kl}^{(t)}$ from $\text{N}_d(b_{kl}, B_{kl}),$ where 
$B_{kl} = \left(\tilde{B}_{0k}^{-1} + \mathfrak{n}_{kl}(\Sigma_{kl}^{-1})^{(t)}\right),$ and $b_{kl} = B_{kl}\left(\tilde{B}_{0k}^{-1} b_{0k}^{(t-1)} + (\Sigma_{kl}^{-1})^{(t)}\mathfrak{n}_{kl}\right)$
\EndFor
\LineComment{See Appendix \ref{appendix:mcmc} \ref{item:hyperpara}}
\For{$j = 1$ to $d$}
\State Sample $\lambda_{kj}$ from $\text{GIG}(p_{kL}, a_{kj}, b_{kj}),$
where $b_{kj} = \sum_{l}\left(\mu_{kl,j}^{(t-1)} - b_{0k,j}^{(t-1)}\right)^2/B_{0,jj}$
\EndFor
\State Sample $C_{0k}$ from $\text{W}_d\left(g_0 + Lc_0, G_0 + \sum_{l = 1}^L(\Sigma_{kl}^{-1})^{(t)}\right)$
\State Sample $b_{0k}$ from $\text{N}_d \left(\tilde{m}_k^{(t)}, \tilde{M}_k^{(t)}\right)$  
\EndFor 
\EndFor
\State \Return{$\left\{(\Theta^{(t)}, \eta^{(t)}): t = t_0 + 1, \ldots, \mathfrak{T}\right\}$}, where first $t_0$'s are burn-in iterations
\end{algorithmic}
\end{algorithm}

\section{Definition of clustering validation metrics\label{appendix:metricdef}}
Write true positive, true negative, false positive and false negative as TP, TN, FP and FN respectively. 

The definition of accuracy, F-measure and adjusted rand index is given as follows:
\begin{enumerate}
    \item Accuracy 
    \begin{align*}
    \text{Accuracy} = \frac{\#\text{TP} + \#\text{TN}}{\#\text{observations}}
\end{align*}
\item F-measure \\ 
Precision and Recall are defined as follows:
\begin{align*}
    \text{Precision} = \frac{\#\text{TP}}{\#\text{TP} + \#\text{FP}}, \quad\text{and}\quad \text{Recall} =  \frac{\#\text{TP}}{\#\text{TP} + \#\text{FN}} 
\end{align*}
F-measure is the harmonic mean of precision of recall:
\begin{align*}
     \text{F-measure} = 2 \cdot\frac{\text{precision}\cdot \text{recall}}{\text{precision} + \text{recall}}.
\end{align*}
\item Adjusted rand index (ARI)\\ 
Adjusted rand index is a corrected-for-chance version of Rand index; the definition of Rand index is similar to accuracy. See \cite{hubert1985comparing} for a formal definition of adjusted rand index.
\end{enumerate}

\newpage
%\vskip 0.2in
\bibliography{papers}
\end{document}